\documentclass[onecolumn,draftcls,romanappendices]{IEEEtran}
\usepackage{mathpple}
\usepackage{times}
\usepackage{amsthm,xpatch}
\usepackage{amsmath,amsfonts}
\usepackage{cite}
\usepackage{amssymb}
\usepackage{dsfont}
\usepackage{graphicx, subfigure}
\usepackage{color}
\usepackage{breqn}
\usepackage{mathtools}
\usepackage{bbm}
\usepackage{latexsym}
\usepackage[ruled, linesnumbered]{algorithm2e}
\usepackage{accents}
\usepackage{tikz}
\usepackage[mathscr]{euscript}
\usepackage{bm}
\usepackage[frak=lucida]{mathalpha} 
\usepackage{comment}
\usepackage{rotating}

\newcommand{\myLambda}{\begin{sideways}%
     \begin{sideways}$\mathrm{V}$\end{sideways}\end{sideways}}

\newtheorem{lemma}{Lemma}
\newtheorem{theorem}{Theorem}
\newtheorem{remark}{Remark}

\IEEEoverridecommandlockouts

\begin{document}

\xpatchcmd{\proof}{\hskip\labelsep}{\hskip3.25\labelsep}{}{}

\pagestyle{plain}

\title{\LARGE Single-Server Private Information Retrieval with Side Information Under Arbitrary Popularity Profiles}

\author{Alejandro Gomez-Leos and Anoosheh Heidarzadeh\thanks{The authors are with the Department of Electrical and Computer Engineering, Texas A\&M University, College Station, TX 77843 USA (E-mail: \{alexgomezleos,anoosheh\}@tamu.edu).}
}

%

%


\maketitle 

\thispagestyle{plain}

\begin{abstract}
This paper introduces a generalization of the Private Information Retrieval with Side Information (PIR-SI) problem called Popularity-Aware PIR-SI (PA-PIR-SI). 
The PA-PIR-SI problem includes one or more remote servers storing copies of a dataset of $K$ messages, and a user who knows $M$ out of $K$ messages---the identities of which are unknown to the server---as a prior side information, and wishes to retrieve one of the remaining $K-M$ messages. 
The goal of the user is to minimize the amount of information they must download from the server while revealing no information about the identity of the desired message.
In contrast to PIR-SI, in PA-PIR-SI, the dataset messages are not assumed to be equally popular. 
That is, given the $M$ side information messages, each of the remaining $K-M$ messages is not necessarily equally likely to be the message desired by the user.
In this work, we focus on the single-server setting of PA-PIR-SI, and establish lower and upper bounds on the capacity of this setting---defined as the maximum possible achievable download rate. 
Our upper bound holds for any message popularity profile, and is the same as the capacity of single-server PIR-SI. 
We prove the lower bound by presenting a PA-PIR-SI scheme which takes a novel probabilistic approach---carefully designed based on the popularity profile---to integrate two existing PIR-SI schemes.
The rate of our scheme is strictly higher than that of the only existing PIR-SI scheme applicable to the PA-PIR-SI setting. 
\end{abstract}

\section{Introduction}

In the Private Information Retrieval (PIR) problem, a user wants to obtain one message belonging to a dataset of $K$ messages with copies stored on a single (or multiple) remote server(s), while revealing no information about the identity of the desired message to the server(s). 
The goal of the user is to privately retrieve their desired message while downloading the minimum possible amount of information from the server(s). 
It was shown in~\cite{CGKS1995} that in the single-server setting, the user must download the entire dataset in order to achieve the privacy requirement, whereas in the multi-server setting, the user can achieve a much higher download rate. 
While the maximum achievable download rate---referred to as \emph{capacity}---of single-server PIR was characterized very early on, the capacity of multi-server PIR was left open until the seminal work by Sun and Jafar~\cite{SJ2017}. 

In recent years, several variations of PIR have been studied by the coding and information theory community.
This includes multi-server PIR~\cite{TER2017,TGKHHER2017,SJ2018,BU18,TSC2019,BAWU2020,LTFH2021,SS2021,BAU2021,ZTSP2021ISIT,LJJ2021}, single-server PIR with side information~\cite{KGHERS2017No0,KGHERS2020,HKS2019Journal,HKS2018,HKS2019,KHSO2019,KHSO2021,HS2021,HS2022Reuse,LJ2022}, multi-server PIR with side information~\cite{T2017,WBU2018,WBU2018No2,KKHS12019,KKHS22019,KGHERS2020,CWJ2020,LG2020CISS,KH2021}, multi-message PIR (MPIR)~\cite{BU2018,BU17}, and MPIR with side information~\cite{HKGRS2018,LG2018,SSM2018,HKRS2019,KKHS32019,HS2022LinCap}.



In this work, we revisit the problem of single-server PIR with side information (PIR-SI)~\cite{KGHERS2020}. 
In PIR-SI, the user knows $M$ out of $K$ dataset messages---the identities of which are unknown to the server---as a prior side information, and wants to retrieve one other message without revealing the identity of the desired message to the server. 
As was shown in~\cite{KGHERS2020}, the capacity of single-server PIR-SI is given by $\lceil K/(M+1)\rceil^{-1}$. 
This result hinges on the assumptions that (i) the $M$ side information messages are chosen uniformly at random, and (ii) given these $M$ messages, each of the remaining $K-M$ messages is equally likely to be the message required by the user.
While the assumption~(i) can be readily justified from the server's perspective, the assumption~(ii) may not always be feasible in practice. 
This is because in many real-world scenarios, not all dataset messages are equally popular.
In particular, recent studies show that  
the Zipf, Gamma, or Weibull distributions are more appropriate statistical models for online data access patterns as compared to the uniform distribution~\cite{CKRAM2009,CDL2008,BCFPS99}. 
This implies the need for new PIR models 
which take into account the popularity of the dataset messages. 

In~\cite{VBU2020}, the authors characterize the capacity of PIR under any arbitrary popularity profile.
To the best of our knowledge, there is, however, no prior result on the capacity of PIR-SI under any non-uniform popularity profile. 
Motivated by this, in this work, we introduce a generalization of the PIR-SI problem, referred to as \emph{Popularity-Aware PIR-SI (PA-PIR-SI)}, which takes into account the popularity of the messages. 
In particular, the PA-PIR-SI problem reduces to the PIR-SI problem when all the messages are equally popular.

We focus on the single-server setting of the PA-PIR-SI problem, and for the ease of exposition, we assume that $K$ and $M$ are such that $M+1$ divides $K$. 
We establish lower and upper bounds on the capacity of PA-PIR-SI 
in the single-server setting.
In particular, we show that the capacity is upper bounded by ${(M+1)/K}$. 
Note that this upper bound does not depend on the popularity profile, and is indeed the same as the capacity of PIR-SI under the uniform popularity profile when ${M+1}$ divides $K$. 
To prove the upper bound, we rely on a mix of combinatorial and information-theoretic arguments. 
To derive a lower bound on the capacity, we propose a PA-PIR-SI scheme, referred to as \emph{Randomized Code Selection (RCS)}, which takes into account the message popularity profile.
The RCS scheme takes a novel probabilistic approach---carefully designed based on the  popularity of the messages---for selecting between two existing PIR-SI schemes which were proposed in~\cite{KGHERS2020}.

We present a motivating example that highlights the limitations of the existing PIR-SI schemes under a non-uniform popularity profile, and demonstrates how the RCS scheme can overcome these limitations.
The RCS scheme is applicable for any arbitrary popularity profile, and achieves a rate strictly higher than $1/(K-M)$---which is the rate of the only existing PIR-SI scheme applicable for non-uniform popularity profiles, i.e. the MDS Code scheme of~\cite{KGHERS2020}.
In addition, our simulations for several commonly-used popularity profiles show that when compared to the rate $1/(K-M)$, the rate of the RCS scheme is much closer to the upper bound $(M+1)/K$. 

\section{Problem Setup} \label{sec:formulation}
We denote random variables by bold symbols, and denote a realization of a random variable by a regular symbol. 
For a positive integer $i$, we denote $\{1,2,...,i\}$ by $[i]$. 
Moreover, for two positive integers $1\leq i <j$, we denote $\{i,i+1,...,j\}$ by $[i:j]$. 
For any set $\mathrm{T}$, we denote by $[\mathrm{T}]^{N}$ the set of all $N$-subsets of $\mathrm{T}$, and denote [$\mathrm{T}]^{1}$ by $\mathrm{T}$ for simplicity. 
We denote by $\mathbb{F}_q$ a finite field of order $q$, and denote by $\mathbb{F}_{q}^{n}$ the $n$-dimensional vector space over $\mathbb{F}_q$. 

Consider a server that stores a dataset containing $K$ messages $X_1, X_2,...,X_K$, where $X_i \in \mathbb{F}_{q}^{n}$ for all ${i\in [K]}$. 
We assume that the random variables $\mathbf{X}_1,\dots,\mathbf{X}_K$ are independent and uniformly distributed over $\mathbb{F}_{q}^{n}$.  
Thus, $H(\mathbf{X}_i) = B \triangleq n\log_{2}q$ for all $i\in [K]$. 
For simplicity, we further denote $[K]$ by $\mathcal{K}$, and 
denote $\{X_i: i\in \mathrm{T}\}$ by $X_{\mathrm{T}}$ for every $\mathrm{T}\subseteq \mathcal{K}$.

Consider a user who has prior knowledge of $M$ messages $X_{\mathrm{S}}= \{X_i: i\in \mathrm{S}\}$ for some $1\leq M\leq K-1$ and some $\mathrm{S}\in [\mathcal{K}]^{M}$, and wishes to retrieve a single message $X_{\mathrm{W}}$ for some ${\mathrm{W} \in \mathcal{K}\setminus \mathrm{S}}$.\footnote{We treat $\mathrm{W}$ as a singleton (i.e., a set of size $1$), instead of an element of a set. 
Similarly, for the case of $M=1$, we treat $\mathrm{S}$ as a singleton.} 
We refer to $X_{\mathrm{W}}$ as the \emph{demand message}, $X_{\mathrm{S}}$ as the \emph{side information messages}, $\mathrm{W}$ as the \emph{demand index}, and $\mathrm{S}$ as the \emph{side information index set}.

We assume that $\mathbf{S}$ is distributed uniformly over $[\mathcal{K}]^M$, where $[\mathcal{K}]^M$ is the set of all $M$-subsets of $\mathcal{K}$. 
That is, the probability mass function (PMF) of $\mathbf{S}$ is given by 
\begin{equation} \label{pmf:S}
p_{\mathbf{S}}(\mathrm{S}^{*}) = \frac{1}{{\binom{K}{M}}} \quad \forall \mathrm{S}^{*} \in [\mathcal{K}]^M.
\end{equation}

Unlike the existing work on PIR-SI, in this work we do not assume that the conditional distribution of $\mathbf{W}$ given $\mathbf{S}$ is uniform. 
Instead, we consider a more general setting that subsumes the original setting of PIR-SI in~\cite{KGHERS2020}. 
For each $i\in \mathcal{K}$, we associate a \emph{popularity} $\lambda_i>0$ to the message $X_i$, where $\lambda_i$ is assumed to be constant with respect to $K$ (i.e., admitting new messages to the dataset does not change the popularity of the existing messages). 
For instance, $\lambda_i$ can correspond to the average number of times that the message $X_i$ is requested in a day, week, or month.
Without loss of generality, we assume that ${\lambda_1 \geq \lambda_2 \geq ... \geq \lambda_K}$.
We denote the tuple ${(\lambda_1, ... ,\lambda_K)}$ by $\myLambda$, and refer to $\myLambda$ as the \emph{(message) popularity profile}.
We also assume that $\myLambda$ is known by both the user and the server. 
Note that~\cite{KGHERS2020} considers the special case of uniform popularity profile, 
i.e., ${\lambda_1=\lambda_2=\dots=\lambda_K}$. 
For simplicity, we denote 
$\sum_{i \in \mathcal{K}\setminus \mathrm{T}}\lambda_i$ by 
$\lambda_{\overline{\mathrm{T}}}$ for any $\mathrm{T} \subseteq \mathcal{K}$. 

Given a popularity profile $\myLambda$, the conditional PMF of $\mathbf{W}$ given $\mathbf{S}$ in defined as
\begin{equation} \label{pmf:WgiveS}
p_{\mathbf{W} \vert \mathbf{S}}(\mathrm{W}^{*}|\mathrm{S}^{*}) = 
\begin{cases}
    \frac{\lambda_{\mathrm{W}^{*}}}{\lambda_{\overline{\mathrm{S}}^{*}}} & \forall\mathrm{W}^{*} \in \mathcal{K}, \forall\mathrm{S}^{*}\in [\mathcal{K}\setminus \mathrm{W}^{*}]^{M}, \\
    0 & \text{otherwise},
\end{cases}
\end{equation}
where $[\mathcal{K}\setminus \mathrm{W}^{*}]^M$ is the set of all $M$-subsets of ${\mathcal{K}\setminus \mathrm{W}^{*}}$. Note that for fixed $\mathrm{S}^{*}$, $\mathbf{W}$ can realize any index $\mathrm{W}^{*}$ in ${\mathcal{K}\setminus\mathrm{S}^{*}}$, and 
the greater is the popularity $\lambda_{\mathrm{W}^{*}}$, the higher is the probability of $\mathbf{W} = \mathrm{W}^{*}$.
By the chain rule of probability, the joint PMF of $\mathbf{W}$ and $\mathbf{S}$ is given by 
\begin{align}
& p_{\mathbf{W},\mathbf{S}}(\mathrm{W}^{*},\mathrm{S}^{*}) \nonumber \\
& \quad = 
\begin{cases}
\frac{1}{\binom{K}{M}}\frac{\lambda_{\mathrm{W}^{*}}}{\lambda_{\overline{\mathrm{S}}^{*}}} & \forall\mathrm{W}^{*} \in \mathcal{K}, \forall\mathrm{S}^{*}\in [\mathcal{K}\setminus \mathrm{W}^{*}]^{M},\\
0 & \text{otherwise}.
\end{cases} \label{eq:JointPMF}
\end{align}
By marginalizing the joint PMF, 
\begin{equation} \label{pmf:W}
p_{\mathbf{W}}(\mathrm{W}^{*}) = 
\frac{1}{{\binom{K}{M}}}\sum\limits_{\mathrm{S}^{*} \in [\mathcal{K}\setminus \mathrm{W}^{*}]^M} \frac{\lambda_{\mathrm{W}^{*}}}{\lambda_{\overline{S}^{*}}} \quad \forall \mathrm{W}^{*} \in \mathcal{K}.
\end{equation} 
We assume that the joint distribution of $\mathbf{W}$ and $\mathbf{S}$ is known to both the user and the server, whereas the realizations $\mathrm{W}$ and $\mathrm{S}$ are known only by the user and not the server.  

Given the demand index ${\mathrm{W}}$ and the side information index set ${\mathrm{S}}$, the user sends a query $\mathrm{Q}^{[\mathrm{W},\mathrm{S}]}$ which is a (potentially stochastic) function of $\mathrm{W}$ and $\mathrm{S}$. 
The server responds with an answer $\mathrm{A}^{[\mathrm{W},\mathrm{S}]}$ which is a deterministic function of 
the user's query $\mathrm{Q}^{[\mathrm{W},\mathrm{S}]}$ and the messages $X_1,\dots,X_K$. 
That is, 
\begin{equation} \label{cond:wellbehave}
    H(\mathbf{A}^{[\mathrm{W},\mathrm{S}]} \vert \mathbf{Q}^{[\mathrm{W},\mathrm{S}]},\mathbf{X}_{\mathcal{K}}) = 0. 
\end{equation}
The randomness in $\mathbf{Q}^{[\mathrm{W},\mathrm{S}]}$ is due to the (potential) randomness in the query construction, and the randomness in $\mathbf{A}^{[\mathrm{W},\mathrm{S}]}$ is due to the (potential) randomness in $\mathbf{Q}^{[\mathrm{W},\mathrm{S}]}$ and the randomness in $\mathbf{X}_{\mathcal{K}}$. 
When there is no danger of confusion, we denote $\mathbf{Q}^{[\mathrm{W},\mathrm{S}]}$, $\mathbf{A}^{[\mathrm{W},\mathrm{S}]}$, $\mathrm{Q}^{[\mathrm{W},\mathrm{S}]}$, and $\mathrm{A}^{[\mathrm{W},\mathrm{S}]}$ by $\mathbf{Q}$, $\mathbf{A}$, $\mathrm{Q}$, and $\mathrm{A}$, respectively. 
We require that the query $\mathrm{Q}$ and the answer $\mathrm{A}$ satisfy the following two conditions:  

\begin{enumerate}
\item \emph{Decodability:} Given $\mathrm{Q}$ and $X_{\mathrm{S}}$, the user must be able to decode the demand $X_{\mathrm{W}}$ from $\mathrm{A}$,
i.e., 
\begin{equation*} \label{cond:decode}
    H(\mathbf{X}_{\mathrm{W}} | \mathbf{A},\mathbf{Q}, \mathbf{X}_{\mathrm{S}}) = 0. 
\end{equation*}
\item \emph{Privacy:} The server must not gain any information about the demand index $\mathrm{W}$ from the query $\mathrm{Q}$, i.e., 
\begin{equation*} \label{cond:privacy}
\mathbb{P}(\mathbf{W}=\mathrm{W}^{*} | \mathbf{Q}= \mathrm{Q}) = \mathbb{P}(\mathbf{W}=\mathrm{W}^{*}) \quad \forall \mathrm{W}^{*}\in \mathcal{K}.
\end{equation*}
\end{enumerate}

Given a popularity profile $\myLambda$, 
the problem is to design a protocol for generating $\mathrm{Q}^{[\mathrm{W},\mathrm{S}]}$ and $\mathrm{A}^{[\mathrm{W},\mathrm{S}]}$ for any realization $(\mathrm{W},\mathrm{S})$ such that both the decodability and privacy conditions are met. 
We refer to this problem as \emph{single-server Popularity-Aware Private Information Retrieval with Side Information (PA-PIR-SI)}.
Since we focus on the single-server setting, we often omit the term ``single-server'' for brevity.

We define the \emph{rate} of a PA-PIR-SI protocol as the ratio of the expected amount of information required by the user, 
i.e., $\sum_{\mathrm{W}^{*}\in \mathcal{K}} p_{\mathbf{W}}(\mathrm{W}^{*})H(\mathbf{X}_{\mathrm{W}^{*}}) = B$, to the expected amount of information downloaded from the server, i.e., $\sum_{\mathrm{W}^{*}\in \mathcal{K}}\sum_{\mathrm{S}^{*}\in [\mathcal{K}\setminus \mathrm{W}^{*}]^{M}} p_{\mathbf{W},\mathbf{S}}(\mathrm{W}^{*},\mathrm{S}^{*})H(\mathbf{A}^{[\mathrm{W}^{*},\mathrm{S}^{*}]})$. 
For a given popularity profile $\myLambda$, we define the \emph{capacity} of PA-PIR-SI as the supremum of rates over all PA-PIR-SI protocols for the popularity profile $\myLambda$. 

Our goal is to derive tight lower and upper bounds on the capacity of PA-PIR-SI for any arbitrary popularity profile.



\section{A Motivating Example}\label{sec:ex}
In this section, we present a motivating example. 
Through this example,
we first overview the existing PIR-SI schemes under the uniform popularity profile, and 
highlight the limitations of these schemes under a non-uniform popularity profile. 
Next, we build upon these schemes and propose a popularity-aware PIR-SI scheme that overcomes the limitations of the PIR-SI schemes that are designed under the uniform popularity profile assumption.   

Consider a server that stores the messages $X_1,\dots,X_6$, and 
a user who knows the message $X_2$ as a prior side information (i.e., $\mathrm{S} = \{2\}$) and 
wishes to retrieve the message $X_1$ (i.e., $\mathrm{W} = \{1\}$). 
Note that in this example, $K=6$ and $M=1$. 
We consider two scenarios for the popularity profile $\myLambda$: (i) ${\lambda_1=\lambda_2=\dots=\lambda_6}$, and (ii) ${\lambda_1=2\lambda_2=\dots=2\lambda_6}$. 
Note that in the case (i), all messages $X_1,\dots,X_6$ are equally popular, whereas in the case (ii), the message $X_1$ is twice more popular than each of the rest of the messages $X_2,\dots,X_6$. 

First, consider the case (i). 
The user can follow the MDS Code scheme of~\cite{KGHERS2020}, and request ${K-M = 5}$ coded combinations of ${X_1,\dots,X_6}$ from the server, 
where the coefficient vectors corresponding to these coded combinations form the rows of the generator matrix of a ${[K=6,K-M=5]}$ MDS code. 
Upon receiving these MDS-coded combinations from the server, the user subtracts off the contribution of $X_2$ from each of these $5$ coded combinations, and obtains $5$ coded combinations of ${X_1,X_3,\dots,X_6}$. 
Since the coefficient vectors pertaining to the resulting coded combinations are linearly independent (by the properties of MDS codes), the user can decode $X_1$ (and ${X_3,\dots,X_6}$) by solving a system of $5$ linear equations with $5$ unknowns ${X_1,X_3,\dots,X_6}$. 
Hence, this scheme satisfies the decodability condition. 
In addition, this scheme naturally satisfies the privacy condition because the user's query is the same for all realizations $(\mathrm{W},\mathrm{S})$. 

Note that for this example, the rate of the MDS Code scheme is ${1/(K-M) = 1/5}$. 
This rate, however, is not optimal. 
As shown in~\cite{HKGRS2018}, the user can follow the Partition-and-Code scheme of~\cite{KGHERS2020} to achieve a higher rate of ${\lceil K/(M+1)\rceil^{-1} = 1/3}$. 
To do so, the user randomly partitions the message indices $1,\dots,6$ into $3$ parts each of size $2$, 
such that one part contains both the demand index $1$ and the side information index $2$, 
say, 
the partition ${\{\{1,2\},\{3,5\},\{4,6\}\}}$. 
Then, the user requests the $3$ coded combinations $X_1+X_2$, $X_3+X_5$, and $X_4+X_6$ from the server.
This scheme satisfies the decodability condition because the user can decode $X_1$ by subtracting off $X_2$ from $X_1+X_2$. 
In the following, we show that this scheme also satisfies the privacy condition. 

Since ${\lambda_1=\lambda_2=\dots=\lambda_6}$, it is easy to verify that ${p_{\mathbf{W},\mathbf{S}}(\{i\},\{j\}) = \frac{1}{30}}$ for all ${i\in [6]}$ and all ${j\in [6]\setminus \{i\}}$, and ${p_{\mathbf{W}}(\{i\}) = \frac{1}{6}}$ for all ${i\in [6]}$. 
For instance, \[p_{\mathbf{W},\mathbf{S}}(\{1\},\{2\}) = \frac{1}{6}\times \frac{\lambda_1}{\sum_{i\in [6]\setminus \{2\}}\lambda_i} = \frac{1}{6}\times \frac{\lambda_1}{5\lambda_1} = \frac{1}{30},\] and 
$p_{\mathbf{W}}(\{1\}) = \sum_{j=1}^{6} p_{\mathbf{W},\mathbf{S}}(\{1\},\{j\}) = \frac{1}{6}$, noting that ${p_{\mathbf{W},\mathbf{S}}(\{1\},\{1\})=0}$ and $p_{\mathbf{W},\mathbf{S}}(\{1\},\{2\})=\dots = p_{\mathbf{W},\mathbf{S}}(\{1\},\{6\})= \frac{1}{30}$. 
Recall that the user's query is given by $\mathrm{Q}= \{\{1,2\},\{3,5\},\{4,6\}\}$. 
To verify that the privacy condition is satisfied, 
we need to show that ${\mathbb{P}(\mathbf{W}=\{i\}|\mathbf{Q}=\mathrm{Q}) = \mathbb{P}(\mathbf{W}=\{i\})}$ for all $i\in [6]$. 
Consider the case of $i=1$ as an example. 
We can write
\begin{align*}
& \mathbb{P}(\mathbf{W}=\{1\}|\mathbf{Q}=\mathrm{Q}) \\
& \quad \stackrel{\scriptsize{\text{(a)}}}{=} \mathbb{P}(\mathbf{W}=\{1\},\mathbf{S}=\{2\}|\mathbf{Q}=\mathrm{Q}) \\
& \quad \stackrel{\scriptsize{\text{(b)}}}{=} \frac{\mathbb{P}(\mathbf{Q}=\mathrm{Q}|\mathbf{W}=\{1\},\mathbf{S}=\{2\})p_{\mathbf{W},\mathbf{S}}(\{1\},\{2\})}{\mathbb{P}(\mathbf{Q}=\mathrm{Q})}
\end{align*}
where (a) follows because $\{1,2\}$ is one of the $3$ parts in the partition $\mathrm{Q} = \{\{1,2\},\{3,5\},\{4,6\}\}$, and hence, if $\mathbf{W}=\{1\}$ (or $\mathbf{W}=\{2\}$), then $\mathbf{S}=\{2\}$ (or $\mathbf{S}=\{1\}$), and (b) follows from Bayes' rule. 

Recall that $p_{\mathbf{W},\mathbf{S}}(\{1\},\{2\}) = \frac{1}{30}$. 
It is also easy to see that ${\mathbb{P}(\mathbf{Q}=\mathrm{Q}|\mathbf{W}=\{1\},\mathbf{S}=\{2\}) = \frac{1}{3}}$. 
This is because given that ${\mathbf{W}=\{1\}}$ and ${\mathbf{S}=\{2\}}$, 
one of the $3$ parts must be $\{1,2\}$, and 
there are $3$ ways to partition $\{3,4,5,6\}$ into $2$ parts each of size $2$, and hence, $\mathbf{Q}$ is equally likely to be either of the $3$ partitions: $\{\{1,2\},\{3,4\},\{5,6\}\}$, $\{\{1,2\},\{3,5\},\{4,6\}\}$, or $\{\{1,2\},\{3,6\},\{4,5\}\}$. 
More generally, it can be seen that \[{\mathbb{P}(\mathbf{Q}=\mathrm{Q}|\mathbf{W}=\{i\},\mathbf{S}=\{j\}) = \frac{1}{3}}\] for all ${(i,j)\in \{(1,2),(2,1),(3,5),(5,3),(4,6),(6,4)\}}$, and \[{\mathbb{P}(\mathbf{Q}=\mathrm{Q}|\mathbf{W}=\{i\},\mathbf{S}=\{j\}) = 0}\] otherwise.
By the total probability theorem, it then follows that \[\mathbb{P}(\mathbf{Q}=\mathrm{Q}) = {\sum_{i,j\in [6]} \mathbb{P}(\mathbf{Q}=\mathrm{Q}|\mathbf{W}=\{i\},\mathbf{S}=\{j\})p_{\mathbf{W},\mathbf{S}}(\{i\},\{j\})= \frac{1}{15}}.\]
Combining these results, we have \[\mathbb{P}(\mathbf{W}=\{1\}|\mathbf{Q}=\mathrm{Q})=\frac{\frac{1}{3}\times \frac{1}{30}}{\frac{1}{15}} = \frac{1}{6}.\] 
Similarly, it can be shown that  $\mathbb{P}(\mathbf{W}=\{i\}|\mathbf{Q}=\mathrm{Q}) = \frac{1}{6}$ for all ${i\in [6]}$. 
Recall that $\mathbb{P}(\mathbf{W}=\{i\}) = p_{\mathbf{W}}(\{i\}) = \frac{1}{6}$ for all ${i\in [6]}$. 
This readily implies that ${\mathbb{P}(\mathbf{W}=\{i\}|\mathbf{Q}=\mathrm{Q})} = {\mathbb{P}(\mathbf{W}=\{i\})}$ for all ${i\in [6]}$, and 
hence, the privacy condition is met.
It should also be noted that by the results of~\cite{HKGRS2018}, the rate $1/3$ is optimal for this example.  

Next, consider the case (ii). 
Recall that, in this case, ${\lambda_1=2\lambda_2=\dots=2\lambda_6}$.
Following the MDS Code scheme as in the case (i), the user requests $5$ MDS-coded combinations of $X_1,\dots,X_6$. 
By using the same arguments as in the case (i), it can be shown that the MDS Code scheme also satisfies the decodability and privacy conditions in the case (ii), and the rate of this scheme is $1/5$ for this example. 
A natural question that arises is whether one can use the Partition-and-Code scheme---similarly as in the case (i)---to achieve a higher rate than $1/5$ in the case (ii). 
We answer this question in the negative, and show that the Partition-and-Code scheme does not always satisfy the privacy condition under a non-uniform popularity profile.

Suppose that the user follows the Partition-and-Code scheme, and constructs the query (partition) \[\mathrm{Q} =  \{\{1,2\},\{3,5\},\{4,6\}\}.\] 
Since ${\lambda_1=2\lambda_2=\dots=2\lambda_6}$, it is easy to verify that  $p_{\mathbf{W},\mathbf{S}}(\cdot,\cdot)$ is given as follows:
\begin{equation*}
    p_{\mathbf{W},\mathbf{S}}(\{i\},\{j\}) =
    \begin{cases}
     \frac{1}{18} & i = 1, j\neq 1,\\
     \frac{1}{30} & i\neq 1, j = 1, \\
    \frac{1}{36} & i \neq 1, j \neq 1, i\neq j\\
    0 & i = j.
    \end{cases}
\end{equation*}
\noindent It is also easy to verify that $p_{\mathbf{W}}(\cdot)$ is given as follows:
\begin{equation*}
    p_{\mathbf{W}}(\{i\}) =
    \begin{cases}
     \frac{5}{18} & i = 1,\\
     \frac{13}{90} & i \in [2:6]. \\
    \end{cases}
\end{equation*}
Using the same technique as in the case (i), it can be shown that $\mathbb{P}(\mathbf{W}=\{1\}|\mathbf{Q}=\mathrm{Q}) = \mathbf{P}(\mathbf{W}=\{1\})$. 
This is because
\begin{align*}
& \mathbb{P}(\mathbf{W}=\{1\}|\mathbf{Q}=\mathrm{Q}) \\ 
& \quad = \frac{\mathbb{P}(\mathbf{Q}=\mathrm{Q}|\mathbf{W}=\{1\},\mathbf{S}=\{2\})p_{\mathbf{W},\mathbf{S}}(\{1\},\{2\})}{\mathbb{P}(\mathbf{Q}=\mathrm{Q})} \\
& \quad = \frac{\frac{1}{3}\times \frac{1}{18}}{\frac{1}{3}\times(\frac{1}{18}+\frac{1}{30}+\frac{1}{36}+\frac{1}{36}+\frac{1}{36}+\frac{1}{36})} = \frac{5}{18},
\end{align*} and $\mathbb{P}(\mathbf{W}=\{1\}) = p_{\mathbf{W}}(\{1\}) = \frac{5}{18}$. 
Note, however, that  ${\mathbb{P}(\mathbf{W}=\{2\}|\mathbf{Q}=\mathrm{Q}) \neq \mathbf{P}(\mathbf{W}=\{2\})}$. 
This is because 
\begin{align*}
& \mathbb{P}(\mathbf{W}=\{2\}|\mathbf{Q}=\mathrm{Q}) \\ 
& \quad = \frac{\mathbb{P}(\mathbf{Q}=\mathrm{Q}|\mathbf{W}=\{2\},\mathbf{S}=\{1\})p_{\mathbf{W},\mathbf{S}}(\{2\},\{1\})}{\mathbb{P}(\mathbf{Q}=\mathrm{Q})} \\
& \quad = \frac{\frac{1}{3}\times \frac{1}{30}}{\frac{1}{3}\times(\frac{1}{18}+\frac{1}{30}+\frac{1}{36}+\frac{1}{36}+\frac{1}{36}+\frac{1}{36})} = \frac{1}{6},
\end{align*} and $\mathbb{P}(\mathbf{W}=\{2\}) = p_{\mathbf{W}}(\{2\}) = \frac{13}{90}$. 
This confirms that the privacy condition is violated, and hence, the Partition-and-Code scheme is not applicable for this case. 
Now, the question is whether there exists any popularity-aware PIR-SI scheme that can outperform the MDS Code scheme for this example. 
We answer this question in the affirmative by presenting a scheme that achieves a rate strictly higher than the rate $1/5$ that can be achieved by the MDS Code scheme.\vspace{0.125cm} 

\emph{Proposed Scheme:} In this scheme, the user takes a randomized approach to choose between the Partition-and-Code scheme and the MDS Code scheme. 
Given $\mathrm{W}=\{i\}$ and $\mathrm{S}=\{j\}$, the user follows the Partition-and-Code or MDS Code scheme with probability $\Gamma_{i,j}$ or $1-\Gamma_{i,j}$, respectively, where $\Gamma_{i,j}$'s for all $i,j\in [6]$ are given as follows: 

\begin{equation*}
    \Gamma_{i,j} =
    \begin{cases}
     \frac{25}{26} & i = 1, j\neq 1,\\
     \frac{5}{6} & i\neq 1, j = 1,\\
     1 & i \neq 1, j \neq 1, i\neq j.\\
    \end{cases}
\end{equation*}
Note that $\Gamma_{1,1},\dots,\Gamma_{6,6}$ are not defined because $\mathrm{W}=\{i\}$ and $\mathrm{S}=\{j\}$ cannot be the same.
As will be shown shortly, the rest of the $\Gamma_{i,j}$'s are chosen carefully---depending on the popularity profile in the case (ii)---such that the privacy condition is satisfied.
The decodability condition is also met because both the Partition-and-Code and MDS Code schemes satisfy the decodability condition.

Recall that ${\mathrm{W}=\{1\}}$ and ${\mathrm{S}=\{2\}}$ in our example. 
Thus, the user either constructs their query following the Partition-and-Code scheme with probability ${\Gamma_{1,2} = \frac{25}{26}}$, or they follow the MDS Code scheme for constructing their query with probability ${1-\Gamma_{1,2}=\frac{1}{26}}$.
Recall that the Partition-and-Code scheme results in requesting $3$ coded combinations, whereas the MDS Code scheme results in requesting $5$ coded combinations. 
Since the expected number of requested coded combinations is ${\frac{25}{26}\times 3+\frac{1}{26}\times 5 = \frac{40}{13}}$, the rate of the proposed scheme is $13/40$ (${>1/5}$).

It remains to verify that the proposed scheme satisfies the privacy condition. 
First, suppose that the user chooses the MDS Code scheme. 
In this case, the query construction is independent of the realization $(\mathrm{W},\mathrm{S})$, and hence, it should be obvious that the privacy condition is met. 
Now, suppose that the user chooses the Partition-and-Code scheme, and constructs the query $\mathrm{Q}=\{\{1,2\},\{3,5\},\{4,6\}\}$. 
We need to show that ${\mathbb{P}(\mathbf{W}=\{i\}|\mathbf{Q}=\mathrm{Q}) = \mathbb{P}(\mathbf{W}=\{i\})}$ for all ${i\in [6]}$.
As an example, consider the case of $i=1$.
Similarly as before, we can write
\begin{align*}
& \mathbb{P}(\mathbf{W}=\{1\}|\mathbf{Q}=\mathrm{Q}) \\
& \quad = \frac{\mathbb{P}(\mathbf{Q}=\mathrm{Q}|\mathbf{W}=\{1\},\mathbf{S}=\{2\})p_{\mathbf{W},\mathbf{S}}(\{1\},\{2\})}{\mathbb{P}(\mathbf{Q}=\mathrm{Q})}.
\end{align*}
Recall that in this case,  $p_{\mathbf{W},\mathbf{S}}(\{1\},\{2\}) = \frac{1}{18}$.
It is easy to see that $\mathbb{P}(\mathbf{Q}=\mathrm{Q}|\mathbf{W}=\{1\},\mathbf{S}=\{2\}) = \Gamma_{1,2} \times \frac{1}{3} = \frac{25}{78}$.
This is because $\mathrm{Q}$ is constructed by the Partition-and-Code scheme for $\mathrm{W} = \{1\}$ and $\mathrm{S}=\{2\}$ (hence, with probability $\Gamma_{1,2}$), and as discussed before, there are $3$ ways to partition the remaining indices $\{3,4,5,6\}$ into $2$ parts each of size $2$.
By the total probability theorem and using $p_{\mathbf{W},\mathbf{S}}(\cdot,\cdot)$ for the popularity profile $\lambda$ in the case (ii), we have
\begin{align*}
& \mathbb{P}(\mathbf{Q}=\mathrm{Q}) \\
& \quad = \sum_{i,j \in [6]}\mathbb{P}(\mathbf{Q}=\mathrm{Q}|\mathbf{W}=\{i\},\mathbf{S}=\{j\}) p_{\mathbf{W},\mathbf{S}}(\{i\},\{j\})\\
& \quad = 
\frac{1}{3}\left(\Gamma_{1,2}\times\frac{1}{18}+\Gamma_{2,1}\times\frac{1}{30}+\Gamma_{3,5}\times\frac{1}{36} \right. \\
& \quad \quad \quad \quad +\left. \Gamma_{5,3}\times\frac{1}{36}+\Gamma_{4,6}\times\frac{1}{36}+\Gamma_{6,4}\times\frac{1}{36}\right)  = \frac{5}{78}.
\end{align*}
Combining these results, it follows that
\begin{align*}
& \mathbb{P}(\mathbf{W}=\{1\}|\mathbf{Q}=\mathrm{Q}) = \frac{\frac{25}{78}\times \frac{1}{18}}{\frac{5}{78}} = \frac{5}{18}.
\end{align*}
Recall that in the case (ii), $\mathbb{P}(\mathbf{W}=\{1\}) = p_{\mathbf{W}}(\{1\}) = \frac{5}{18}$. 
Thus, ${\mathbb{P}(\mathbf{W}=\{1\}|\mathbf{Q}=\mathrm{Q})} = \mathbb{P}(\mathbf{W}=\{1\})=\frac{5}{18}$.

Now, let us consider the case of $i=2$ as another example.
Similarly, we have
\begin{align*}
& \mathbb{P}(\mathbf{W}=\{2\}|\mathbf{Q}=\mathrm{Q})\\
& \quad = \frac{\mathbb{P}(\mathbf{Q}=\mathrm{Q}|\mathbf{W}=\{2\},\mathbf{S}=\{1\})p_{\mathbf{W},\mathbf{S}}(\{2\},\{1\})}{\mathbb{P}(\mathbf{Q}=\mathrm{Q})}.
\end{align*}
Recall that $p_{\mathbf{W},\mathbf{S}}(2,1) = \frac{1}{30}$.
By the same arguments as in the previous example, ${\mathbb{P}(\mathbf{Q}=\mathrm{Q}|\mathbf{W}=\{2\},\mathbf{S}=\{1\})} = \Gamma_{2,1} \times \frac{1}{3} = \frac{5}{18}$.
Notice that $\mathrm{Q}$ is the same as in the previous example, and hence, 
${\mathbb{P}(\mathbf{Q}=\mathrm{Q}) = \frac{5}{78}}$, as shown earlier. 
Combining these results, we have
\begin{align*}
& \mathbb{P}(\mathbf{W}=\{2\}|\mathbf{Q}=\mathrm{Q}) = \frac{\frac{5}{18}\times \frac{1}{30}}{\frac{5}{78}} = \frac{13}{90}.
\end{align*}
Recall that $\mathbb{P}(\mathbf{W}=\{2\}) = p_{\mathbf{W}}(\{2\}) = \frac{13}{90}$ in the case (ii). 
Thus, ${\mathbb{P}(\mathbf{W}=\{2\}|\mathbf{Q}=\mathrm{Q})} = \mathbb{P}(\mathbf{W}=\{2\})=\frac{13}{90}$.

Similarly as in the cases of $i=1$ and $i=2$, it can be shown that ${\mathbb{P}(\mathbf{W}=\{i\}|\mathbf{Q}=\mathrm{Q}) = \mathbb{P}(\mathbf{W}=\{i\})}$ for all ${i\in [6]}$. 
This completes the proof of privacy.


\section{Main Results}
In this section, we summarize our main results on the capacity of PA-PIR-SI.
\begin{theorem}\label{thm:1}
For PA-PIR-SI with $K$ messages and $M$ side information messages such that $M+1$ is a divisor of $K$ and strictly less than $\sqrt{K}$, 
under any popularity profile $\myLambda$, 
the capacity is upper bounded by $R_{\emph{\text{UB}}}$ defined as \begin{equation}\label{eq:RUB}
\frac{M+1}{K},    
\end{equation} and is lower bounded by $R_{\emph{\text{LB}}}$ defined as
\begin{align} \label{eq:thm1_rate}
& \hspace{-0.3cm}\left(K-M - \Biggl(K-M-\frac{K}{M+1}\right) \nonumber\\
& \quad \times \Gamma_{\{1\},[2:M+1]}\frac{p_{\mathbf{W},\mathbf{S}}(\{1\},[2:M+1])}{p_{\mathbf{W}}(\{1\})} \binom{K-1}{M}\Biggr)^{-1},
\end{align}
where $\Gamma_{\{1\}, [2:M+1]}$ is given by
\begin{align} \label{eq:thm1_parameter}
    & \min_{i \in [K-M:K]}\Biggl\{1,\frac{p_{\mathbf{W},\mathbf{S}}(\{i\},[K-M:K]\setminus \{i\})p_{\mathbf{W}}(\{1\}) }{p_{\mathbf{W},\mathbf{S}}(\{1\},[2:M+1])p_{\mathbf{W}}(\{i\})}\Biggr\}, 
\end{align}
and $p_{\mathbf{W},\mathbf{S}}(\cdot,\cdot)$ and $p_{\mathbf{W}}(\cdot)$ depend on the popularity profile $\myLambda$, and are defined as in~\eqref{eq:JointPMF} and~\eqref{pmf:W}, respectively.  
\end{theorem}

The proof of converse (i.e., the upper bound on the capacity) is based on information-theoretic arguments. 
The key ingredient in the converse proof is a necessary condition for any PA-PIR-SI protocol due to the decodability and privacy conditions. 
To prove the achievability result 
(i.e., the lower bound on the capacity), 
we build upon the existing PIR-SI schemes under uniform popularity profile, and 
propose a popularity-aware PIR-SI scheme that is applicable to any arbitrary popularity profile. 
The proposed scheme takes a randomized approach---carefully designed based on the popularity profile---for selecting between two different techniques for query construction. 
\begin{remark}\normalfont 
Note that the lower bound $R_{\text{LB}}$---which is the rate achieved by our scheme---is valid only for $K$ and $M$ such that ${(M+1)\mid K}$ and ${M+1< \sqrt{K}}$, whereas the upper bound $R_{\text{UB}}$ holds for all $K$ and $M$. 
While our scheme can be modified so that it is applicable for all $K$ and $M$, the modified scheme's description is lengthy and notation-heavy, and its analysis is tedious and involved. 
To avoid confusing the reader with technical details, in this work we present the simplest form of our scheme (i.e., for $K$ and $M$ satisfying the above conditions), and demonstrate its superiority over the MDS Code scheme of~\cite{KGHERS2020}---which is the only existing PIR-SI scheme applicable for arbitrary popularity profiles.
\end{remark}

\begin{remark}\normalfont 
By the result of~\cite[Theorem~1]{VBU2020} on the capacity of semantic PIR, the capacity of single-server PIR (without side information) under any arbitrary (uniform or non-uniform) popularity profile is $1/K$. 
That is, the privacy can be achieved only by downloading the entire dataset.
The result of Theorem~\ref{thm:1} shows that for any popularity profile, the capacity of single-server PA-PIR-SI is between $1/(K-M)$ and $(M+1)/K$, and hence, greater than $1/K$.
This result extends our prior understanding of the role of side information in single-server PIR-SI under the uniform popularity profile, to arbitrary popularity profiles. 
\end{remark}

\section{Proof of Theorem 1}\label{sec:achieve1}
In this section, we present the converse and achievability proofs for Theorem~\ref{thm:1}. 
The proofs of all lemmas are given in Appendix.  

\subsection{Converse Proof}\label{subsec:Converse}
Fix arbitrary $\mathrm{W}\in \mathcal{K}$ and $\mathrm{S}\in [\mathcal{K}\setminus \mathrm{W}]^{M}$. 
Consider an arbitrary PA-PIR-SI protocol. 
Recall that the rate of a protocol is equal to the ratio of $B$ to $\sum_{\mathrm{W}^{*}\in \mathcal{K}}\sum_{\mathrm{S}^{*}\in [\mathcal{K}\setminus \mathrm{W}^{*}]^{M}} p_{\mathbf{W},\mathbf{S}}(\mathrm{W}^{*},\mathrm{S}^{*})H(\mathbf{A}^{[\mathrm{W}^{*},\mathrm{S}^{*}]})$. 
To prove that the capacity is upper bounded by ${(M+1)/K}$, we need to show that ${H(\mathbf{A}^{[\mathrm{W},\mathrm{S}]})\geq (K/(M+1))B = NB}$, where $N\triangleq K/(M+1)$, and $B$ is the entropy of a message.
This is because 
if $H(\mathbf{A}^{[\mathrm{W},\mathrm{S}]})\geq NB$ for all $(\mathrm{W},\mathrm{S})$, 
then $\sum_{\mathrm{W}^{*}\in \mathcal{K}}\sum_{\mathrm{S}^{*}\in [\mathcal{K}\setminus \mathrm{W}^{*}]^{M}} p_{\mathbf{W},\mathbf{S}}(\mathrm{W}^{*},\mathrm{S}^{*})H(\mathbf{A}^{[\mathrm{W}^{*},\mathrm{S}^{*}]})\geq NB$, and hence, the rate is upper bounded by ${B/(NB)} = {(M+1)/K}$. 
To show that ${H(\mathbf{A}^{[\mathrm{W},\mathrm{S}]})\geq NB}$, we rely on the following necessary condition for any PA-PIR-SI protocol.  

\begin{lemma}\label{lem:NC}
Given any PA-PIR-SI protocol for any arbitrary popularity profile $\myLambda$, for any given ${\mathrm{W}^{*}\in \mathcal{K}}$, there must exist ${\mathrm{S}^{*}\in [\mathcal{K}\setminus \mathrm{W}^{*}]^{M}}$ such that $\mathrm{X}_{\mathrm{W}^{*}}$ can be recovered from the query and the answer given $\mathrm{X}_{\mathrm{S}^{*}}$, 
i.e., 
\[{H(\mathbf{X}_{\mathrm{W}^{*}}|\mathbf{A}^{[\mathrm{W},\mathrm{S}]},\mathbf{Q}^{[\mathrm{W},\mathrm{S}]},\mathbf{X}_{\mathrm{S}^{*}})=0}.\]
\end{lemma}

Let $\mathrm{W}_0= \emptyset$ and $\mathrm{S}_0 = \emptyset$. 
Take arbitrary distinct ${\mathrm{W}_1,\dots,\mathrm{W}_N\in \mathcal{K}}$ such that ${\mathrm{W}_i\in \mathcal{K}\setminus \cup_{j=0}^{i-1} (\mathrm{W}_j\cup \mathrm{S}_j)}$ for each ${i\in [N]}$, 
where ${\mathrm{S}_i\in [\mathcal{K}\setminus \mathrm{W}_i]^{M}}$ for each ${i\in [N]}$ is such that $H(\mathbf{X}_{\mathrm{W}_i}|\mathbf{A}^{[\mathrm{W},\mathrm{S}]},\mathbf{Q}^{[\mathrm{W},\mathrm{S}]},\mathbf{X}_{\mathrm{S}_i}) = 0$. 
The existence of $N$ such pairs $(\mathrm{W}_1,\mathrm{S}_1),\dots,(\mathrm{W}_N,\mathrm{S}_N)$ is guaranteed by the result of Lemma~\ref{lem:NC} and the fact that ${\mathcal{K}\setminus \cup_{j=0}^{i-1} (\mathrm{W}_j\cup \mathrm{S}_j)\neq \emptyset}$ for any ${i\in [N]}$.\footnote{
Note that ${|\cup_{j=0}^{i-1} (\mathrm{W}_j\cup \mathrm{S}_j)|}\leq {(i-1)(M+1)}$, and 
hence, ${|\mathcal{K}\setminus \cup_{j=0}^{i-1} (\mathrm{W}_j\cup \mathrm{S}_j)|}\geq {K - (i-1)(M+1)} = {(N-i+1)(M+1)}\geq {M+1}>0$ for all ${i\in [N]}$.} 
For simplifying the notation, we denote $\cup_{j=0}^{i-1} (\mathrm{W}_j\cup \mathrm{S}_j)$ by $\mathrm{U}_{i}$ for each $i\in [N+1]$. 
Note that $\mathrm{U}_1 = \mathrm{W}_0\cup \mathrm{S}_0 = \emptyset$. 
Also, we denote $\mathbf{Q}^{[\mathrm{W},\mathrm{S}]}$ and $\mathbf{A}^{[\mathrm{W},\mathrm{S}]}$ by $\mathbf{Q}$ and $\mathbf{A}$, respectively. 

\begin{lemma}\label{lem:Claim}
For each $i\in [N]$, it holds that 
\begin{equation}\label{eq:Claim}
H(\mathbf{A}|\mathbf{Q},\mathbf{X}_{\mathrm{U}_{i}}) \geq H(\mathbf{X}_{\mathrm{W}_{i}})+H(\mathbf{A}|\mathbf{Q},\mathbf{X}_{\mathrm{U}_{i+1}}).
\end{equation}
\end{lemma}

By applying Lemma~\ref{lem:Claim} repeatedly ($N$ times), we can write
\begin{align}\label{eq:HALB}
H(\mathbf{A}) & \stackrel{\scriptsize{\text{(a)}}}{\geq} H(\mathbf{A}|\mathbf{Q}) \nonumber \\
& \stackrel{\scriptsize{\text{(b)}}}{=} H(\mathbf{A}|\mathbf{Q},\mathbf{X}_{\mathrm{U}_1}) \nonumber \\
& \stackrel{\scriptsize{\text{(c)}}}{\geq} H(\mathbf{X}_{\mathrm{W}_1}) + H(\mathbf{A}|\mathbf{Q},\mathbf{X}_{\mathrm{U}_2}) \nonumber \\
& \stackrel{\scriptsize{\text{(d)}}}{\geq} H(\mathbf{X}_{\mathrm{W}_1}) + H(\mathbf{X}_{\mathrm{W}_2}) + H(\mathbf{A}|\mathbf{Q},\mathbf{X}_{\mathrm{U}_3}) \nonumber \\[-0.125cm]
& \hspace{0.2cm} \vdots \nonumber\\
& \stackrel{\scriptsize{\text{(e)}}}{\geq} H(\mathbf{X}_{\mathrm{W}_1}) + \dots + H(\mathbf{X}_{\mathrm{W}_N}) + H(\mathbf{A}|\mathbf{Q},\mathbf{X}_{\mathrm{U}_{N+1}}) \nonumber \\
& \stackrel{\scriptsize{\text{(f)}}}{\geq} H(\mathbf{X}_{\mathrm{W}_1}) + \dots + H(\mathbf{X}_{\mathrm{W}_N}), 
\end{align}
where (a) holds since conditioning does not increase the entropy; 
(b) follows because ${\mathrm{U}_1 = \emptyset}$;
(c), (d), and (e) follow from~\eqref{eq:Claim} for the cases of ${i=1}$, ${i=2}$, and ${i=N}$, respectively; 
and (f) follows from the non-negativity of the entropy.

Since ${H(\mathbf{X}_{\mathrm{W}_i}) = B}$ for all ${i\in [N]}$, it then follows from~\eqref{eq:HALB} that $H(\mathbf{A})\geq NB$, as was to be shown. 

\subsection{Achievability Scheme}\label{subsec:Ach}
In this section, we propose a PA-PIR-SI scheme for arbitrary popularity profiles. 
The proposed scheme, which we refer to as the \emph{Randomized Code Selection (RCS) scheme}, extends the scheme we presented in Section~\ref{sec:ex}, and is applicable for any number of dataset messages $K$ and any number of side information messages $M$ such that $M+1$ is a divisor of $K$ and strictly less than $\sqrt{K}$, and any field size $q\geq K$.\vspace{0.125cm}

\textbf{Randomized Code Selection (RCS) Scheme:} 
For any ${\mathrm{W}^{*}\in \mathcal{K}}$ and ${\mathrm{S}^{*}\in [\mathcal{K}\setminus \mathrm{W}^{*}]^{M}}$, we define 
\begin{equation} \label{eq:GammaWSFunc}
    \Gamma_{\mathrm{W}^{*},\mathrm{S}^{*}} = \Gamma_{\{1\}, [2:M+1]} \frac{p_{\mathbf{W},\mathbf{S}}(\{1\},[2:M+1])p_{\mathbf{W}}(\mathrm{W}^{*})}{p_{\mathbf{W},\mathbf{S}}(\mathrm{W}^{*},\mathrm{S}^{*})p_{\mathbf{W}}(\{1\})}, 
\end{equation}
where $\Gamma_{\{1\}, [2:M+1]}$ is given by~\eqref{eq:thm1_parameter}. 
Given the demand index ${\mathrm{W}}$ and the side information index set ${\mathrm{S}}$, 
the user randomly selects the Partition-and-Code scheme with probability $\Gamma_{\mathrm{W},\mathrm{S}}$, or the MDS Code scheme with probability $1-\Gamma_{\mathrm{W},\mathrm{S}}$, and 
follows the selected scheme as described below.
In the following, we refer to the Partition-and-Code scheme as Scheme~I, and 
refer to the MDS Code scheme as Scheme~II.\vspace{0.125cm} 

{\bf Scheme~I:}  
This scheme consists of the three steps outlined below: 

\emph{Step 1:} The user partitions the message indices $1,\dots,K$ into $N\triangleq K/(M+1)$ parts $\mathrm{Q}_1,\dots,\mathrm{Q}_{N}$, each of size $M+1$, as outlined below. 
First, the user chooses an index $j^{*}\in [N]$ uniformly at random, and assigns the demand index $\mathrm{W}$ and the side information indices $\mathrm{S}$ to the part $\mathrm{Q}_{j^{*}}$. 
The user then takes the remaining $K-(M+1)$ message indices ${\mathcal{K}\setminus (\mathrm{W}\cup\mathrm{S})}$, and randomly partitions them into the remaining $N-1$ parts $\mathrm{Q}_j$'s for $j\in [N]\setminus \{j^{*}\}$.
Then, the user constructs the query $\mathrm{Q}^{[\mathrm{W},\mathrm{S}]} = \{\mathrm{Q}_1,\dots,\mathrm{Q}_N\}$, and sends it to the server. 

\emph{Step 2:} Given $\mathrm{Q}^{[\mathrm{W},\mathrm{S}]}$, the server computes ${\mathrm{A}_j = \sum_{i \in \mathrm{Q}_j}X_i}$ for each ${j\in [N]}$.
Then, the server constructs the answer $\mathrm{A}^{[\mathrm{W},\mathrm{S}]} = \{\mathrm{A}_1,\dots,\mathrm{A}_N\}$, and sends it back to the user. 

\textit{Step 3:} Given $\mathrm{A}^{[\mathrm{W},\mathrm{S}]}$, the user recovers their demand message $X_{\mathrm{W}}$ by subtracting off the contribution of the side information messages $X_{\mathrm{S}}$ from $\mathrm{A}_{j^{*}}$, i.e., 
${X_{\mathrm{W}} = \mathrm{A}_{j^{*}} - \sum_{i\in \mathrm{S}} X_{i}}$.\vspace{0.125cm} 

{\bf Scheme~II:} 
This scheme consists of the following three steps:

\emph{Step 1:} First, the user chooses $K$ arbitrary (but distinct) elements $\omega_1,\dots,\omega_{K}$ from $\mathbb{F}_q$. 
The user then constructs ${K-M}$ vectors $\mathrm{Q}_1,\dots,\mathrm{Q}_{K-M}$, where $\mathrm{Q}_j = [\omega_1^{j-1},...,\omega_{K}^{j-1}]$ for each $j\in[K-M]$. 
Then, the user constructs the query $\mathrm{Q}^{[\mathrm{W},\mathrm{S}]} = \{{\mathrm{Q}_1,\dots,\mathrm{Q}_{K-M}}\}$, and sends it to the server. 

\textit{Step 2:} Given $\mathrm{Q}^{[\mathrm{W},\mathrm{S}]}$, the server computes ${\mathrm{A}_j = \sum_{i=1}^{K}\omega_i^{j-1}X_i}$ for each ${j\in [K-M]}$. 
The server then constructs the answer $\mathrm{A}^{[\mathrm{W},\mathrm{S}]} = \{\mathrm{A}_1,\dots,\mathrm{A}_{K-M}\}$, and sends it back to the user.

\textit{Step 3:} Given $\mathrm{A}^{[\mathrm{W},\mathrm{S}]}$, the user recovers their demand message $X_{\mathrm{W}}$---along with all $K-(M+1)$ messages $X_{\mathcal{K}\setminus (\mathrm{W}\cup\mathrm{S})}$---by subtracting off the contribution of the side information messages $X_{\mathrm{S}}$ from $\mathrm{A}_1,\dots,\mathrm{A}_{K-M}$, and solving the resulting system of $K-M$ linear equations with $K-M$ unknowns $X_{\mathcal{K}\setminus \mathrm{S}}$. 

\subsection{Proof of Decodability and Privacy}\label{subsec:DecPrivacy}
Since both Schemes I and II satisfy the decodability condition, it should be obvious that the RCS scheme also satisfies this requirement. 
It thus remains to show that the RCS scheme also satisfies the privacy condition. 

Consider a query constructed by the RCS scheme. 
When the query is formed by Scheme~II, it should be obvious that the privacy condition is satisfied 
because Scheme~II constructs the query independently of the realization $(\mathrm{W},\mathrm{S})$.
In the following, we show that the privacy condition is also satisfied when the query is formed by Scheme~I. 

Recall that any query formed by Scheme~I is a partition of $\mathcal{K}$ with ${N = K/(M+1)}$ parts, each of size $M+1$.
We denote by $\mathcal{Q}$ the set of all such partitions.
For each $\mathrm{Q}\in \mathcal{Q}$, 
let $\mathrm{Q}_1,\dots,\mathrm{Q}_N$ denote the $N$ parts forming the partition $\mathrm{Q}$. 

\begin{lemma}\label{lem:NCQ}
For any query (partition) $\mathrm{Q}\in \mathcal{Q}$, the privacy condition is satisfied if for any $i,j\in [N]$ and for any ${\mathrm{W}_i\in \mathrm{Q}_i},{\mathrm{W}_j\in \mathrm{Q}_j}$, it holds that
\begin{equation} \label{eq:parameterize}
 \Gamma_{\mathrm{W}_j,\mathrm{S}_j}= \Gamma_{\mathrm{W}_{i},\mathrm{S}_{i}} \frac{p_{\mathbf{W},\mathbf{S}}(\mathrm{W}_{i},\mathrm{S}_{i})p_{\mathbf{W}}(\mathrm{W}_j) }{p_{\mathbf{W},\mathbf{S}}(\mathrm{W}_j,\mathrm{S}_j)p_{\mathbf{W}}(\mathrm{W}_{i})},
\end{equation} where ${\mathrm{S}_i = \mathrm{Q}_i\setminus \mathrm{W}_i}$ and ${\mathrm{S}_j = \mathrm{Q}_j\setminus \mathrm{W}_j}$. 
\end{lemma}

By Lemma~\ref{lem:NCQ}, the privacy requirement entails that the condition in~\eqref{eq:parameterize} must hold for any two parts $\mathrm{Q}_i$ and $\mathrm{Q}_j$ in any partition $\mathrm{Q}$. 
To complete the proof of privacy, 
it thus suffices to show that our choice of $\Gamma_{\mathrm{W}^{*},\mathrm{S}^{*}}$ in the RCS scheme satisfies the condition in~\eqref{eq:parameterize}. 

Fix arbitrary $\mathrm{W}^{*}\in \mathcal{K}$ and $\mathrm{S}^{*}\in [\mathcal{K}\setminus \mathrm{W}^{*}]^{M}$ such that $\mathrm{W}^{*}\cup\mathrm{S}^{*}$ is one of the parts in the partition $\mathrm{Q}$.
We need to show that  $\Gamma_{\mathrm{W}^{*},\mathrm{S}^{*}}$ given by~\eqref{eq:GammaWSFunc} satisfies the condition in~\eqref{eq:parameterize}.
We consider the following cases separately: 
(i) ${\mathrm{Q}_i = [M+1]}$ for some ${i\in [N]}$, and 
(ii) ${\mathrm{Q}_i\neq [M+1]}$ for any ${i\in [N]}$. 

First, consider the case (i). 
Taking ${\mathrm{W}_i = \{1\}}$ and ${\mathrm{S}_i = [2:M+1]}$, the condition in~\eqref{eq:parameterize} reduces to 
\begin{equation*} \label{eq:parameterize2}
\Gamma_{\mathrm{W}^{*},\mathrm{S}^{*}}= \Gamma_{\{1\},[2:M+1]} \frac{p_{\mathbf{W},\mathbf{S}}(\{1\},[2:M+1])p_{\mathbf{W}}(\mathrm{W}^{*}) }{p_{\mathbf{W},\mathbf{S}}(\mathrm{W}^{*},\mathrm{S}^{*})p_{\mathbf{W}}(\{1\})},
\end{equation*} 
which is consistent with our choice of $\Gamma_{\mathrm{W}^{*},\mathrm{S}^{*}}$ (cf.~\eqref{eq:GammaWSFunc}). 

Next, consider the case (ii). 
Recall that by assumption, ${M+1<\sqrt{K}}$, i.e., ${K = N(M+1)>(M+1)^2}$, or equivalently, ${N> M+1}$. 
Since $\mathrm{Q}$ consists of $N>M+1$ parts, and $|[M+1]| = M+1$, by the pigeonhole principle, there exists some $k\in [N]$ such that $\mathrm{Q}_k$ and $[M+1]$ are disjoint. 
Let ${\mathrm{Q}^{*}\in \mathcal{Q}}$ be an arbitrary partition such that both parts ${\mathrm{Q}_{k}}$ and ${[M+1]}$ belong to the partition $\mathrm{Q}^{*}$.
Recall that the privacy condition requires that for any given partition, the condition in~\eqref{eq:parameterize} must hold for any two parts of that partition.
Note that ${\mathrm{Q}_{k}}$ and ${[M+1]}$ are two parts of the same partition $\mathrm{Q}^{*}$. 
Let $\mathrm{W}_{k}$ be an arbitrary index in the part $\mathrm{Q}_{k}$, and let $\mathrm{S}_{k} = \mathrm{Q}_{k}\setminus \mathrm{W}_{k}$.
Then, by~\eqref{eq:parameterize}, it is required that 
\begin{equation} \label{eq:parameterize3}
 \Gamma_{\mathrm{W}_{k},\mathrm{S}_{k}}= \Gamma_{\{1\},[2:M+1]} \frac{p_{\mathbf{W},\mathbf{S}}(\{1\},[2:M+1])p_{\mathbf{W}}(\mathrm{W}_{k}) }{p_{\mathbf{W},\mathbf{S}}(\mathrm{W}_{k},\mathrm{S}_{k})p_{\mathbf{W}}(\{1\})}.
\end{equation}
Note also that $\mathrm{W}^{*}\cup\mathrm{S}^{*}$ and  ${\mathrm{W}_{k}\cup\mathrm{S}_{k}}$ are two parts of the partition $\mathrm{Q}$. 
Thus, by~\eqref{eq:parameterize}, we require that
\begin{equation} \label{eq:parameterize4}
 \Gamma_{\mathrm{W}^{*},\mathrm{S}^{*}}= \Gamma_{\mathrm{W}_k,\mathrm{S}_k} \frac{p_{\mathbf{W},\mathbf{S}}(\mathrm{W}_k,\mathrm{S}_k)p_{\mathbf{W}}(\mathrm{W}^{*}) }{p_{\mathbf{W},\mathbf{S}}(\mathrm{W}^{*},\mathrm{S}^{*})p_{\mathbf{W}}(\mathrm{W}_k)}.
\end{equation}
Combining~\eqref{eq:parameterize3} and~\eqref{eq:parameterize4}, it follows that we must have
\begin{equation*} \label{eq:parameterize5}
 \Gamma_{\mathrm{W}^{*},\mathrm{S}^{*}}= \Gamma_{\{1\},[2:M+1]} \frac{p_{\mathbf{W},\mathbf{S}}(\{1\},[2:M+1])p_{\mathbf{W}}(\mathrm{W}^{*}) }{p_{\mathbf{W},\mathbf{S}}(\mathrm{W}^{*},\mathrm{S}^{*})p_{\mathbf{W}}(\{1\})},
\end{equation*}
which coincides with our choice of $\Gamma_{\mathrm{W}^{*},\mathrm{S}^{*}}$ (cf.~\eqref{eq:GammaWSFunc}).
This completes the proof of privacy.

\subsection{Proof of Achievable Rate}\label{subsec:Rate}
By construction, the server's answer to the user's query consists of $K/(M+1)$ (or ${K-M}$) linearly independent combinations of the messages $X_1,\dots,X_K$ for Scheme~I (or Scheme~II). 
Since $\mathbf{X}_1,\dots,\mathbf{X}_{K}$ are independent and uniformly distributed over $\mathbb{F}_q^n$ (by assumption), 
then $\mathbf{A}_1,\dots,\mathbf{A}_N$ (or $\mathbf{A}_1,\dots,\mathbf{A}_{K-M}$) are independent and uniformly distributed over $\mathbb{F}_q^n$. 
Thus, $H(\mathbf{A}^{[\mathrm{W},\mathrm{S}]})$ is equal to ${H(\mathbf{A}_1,\dots,\mathbf{A}_N)}={NB} = (K/(M+1))B$ (or ${H(\mathbf{A}_1,\dots,\mathbf{A}_{K-M})}={(K-M)B}$) for Scheme~I (or Scheme~II). 
Using the joint PMF of $(\mathbf{W},\mathbf{S})$, it then follows that the rate of the RCS scheme is given by
\begin{align} \label{eq:expect_rate}
    & \biggl(\sum_{\mathrm{W}^{*}\in \mathcal{K},\mathrm{S}^{*}\in [\mathcal{K}\setminus \mathrm{W}^{*}]^{M}} p_{\mathbf{W},\mathbf{S}}(\mathrm{W}^{*},\mathrm{S}^{*}) \nonumber \\ 
    & \quad\quad \times  {\left[\Gamma_{\mathrm{W}^{*},\mathrm{S}^{*}}\left(\frac{K}{M+1}\right)+(1-\Gamma_{\mathrm{W}^{*},\mathrm{S}^{*}})\left(K-M\right)\right]} \biggr)^{-1}.
\end{align} 
Substituting for $\Gamma_{\mathrm{W}^{*},\mathrm{S}^{*}}$ as defined in~\eqref{eq:GammaWSFunc}, it is easy to verify that~\eqref{eq:expect_rate} reduces to
\begin{align} \label{eq:reduced_rate}
    & \left(K-M - \left(K-M-\frac{K}{M+1}\right) \right. \nonumber \\
    & \quad\quad \left. \times \Gamma_{\{1\},[2:M+1]}\frac{p_{\mathbf{W},\mathbf{S}}(\{1\},[2:M+1])}{p_{\mathbf{W}}(\{1\})} \binom{K-1}{M}\right)^{-1},
\end{align} 
which is the same as the expression for $R_{\text{LB}}$ in Theorem~\ref{thm:1} (cf.~\eqref{eq:thm1_rate}).

Since $\Gamma_{\mathrm{W}^{*},\mathrm{S}^{*}}$'s are probabilities, they can only take values in the interval $[0,1]$, 
i.e., for any ${\mathrm{W}^{*}\in \mathcal{K}}$ and any ${\mathrm{S}^{*}\in [\mathcal{K}\setminus \mathrm{W}^{*}]^{M}}$, it must hold that
\begin{equation*}
    0\leq \Gamma_{\{1\}, [2:M+1]} \frac{p_{\mathbf{W},\mathbf{S}}(\{1\},[2:M+1])p_{\mathbf{W}}(\mathrm{W}^{*})}{p_{\mathbf{W},\mathbf{S}}(\mathrm{W}^{*},\mathrm{S}^{*})p_{\mathbf{W}}(\{1\})}\leq 1, 
\end{equation*} 
which implies that $\Gamma_{\{1\},[2:M+1]}$ is lower bounded by $0$, and upper bounded by  
\begin{equation} \label{ineq:param}
   \min_{\mathrm{W}^{*}, \mathrm{S}^{*}}\left\{1,\frac{p_{\mathbf{W},\mathbf{S}}(\mathrm{W}^{*},\mathrm{S}^{*})p_{\mathbf{W}}(\{1\}) }{p_{\mathbf{W},\mathbf{S}}(\{1\},[2:M+1])p_{\mathbf{W}}(\mathrm{W}^{*})}\right\},
\end{equation} 
where the minimization is over all ${\mathrm{W}^{*}\in \mathcal{K}}$ and all ${\mathrm{S}^{*}\in [\mathcal{K}\setminus \mathrm{W}^{*}]^{M}}$. 
According to~\eqref{eq:reduced_rate}, for fixed $K$ and $M$, the rate of the RCS scheme is an increasing function of $\Gamma_{\{1\},[2:M+1]}$, and 
hence, the rate is maximized when $\Gamma_{\{1\},[2:M+1]}$ is equal to~\eqref{ineq:param}. 
It remains to show that~\eqref{ineq:param} and our choice of $\Gamma_{\{1\},[2:M+1]}$ given by~\eqref{eq:thm1_parameter} are equal.   

It is more convenient to analyze the following minimization problem (instead of the one in~\eqref{ineq:param}): 
\begin{equation} \label{eq:search_reduced}
   \min_{\mathrm{W}^{*},\mathrm{S}^{*}}\left\{\frac{p_{\mathbf{W},\mathbf{S}}(\{1\},[2:M+1])}{p_{\mathbf{W}}(\{1\})},\frac{p_{\mathbf{W},\mathbf{S}}(\mathrm{W}^{*},\mathrm{S}^{*})}{p_{\mathbf{W}}(\mathrm{W}^{*})}\right\},
\end{equation} 
where the minimization is over all ${\mathrm{W}^{*}\in \mathcal{K}}$ and all ${\mathrm{S}^{*}\in [\mathcal{K}\setminus \mathrm{W}^{*}]^{M}}$.
Note that~\eqref{ineq:param} is equal to~\eqref{eq:search_reduced} times the constant term $p_{\mathbf{W}}(\{1\})/p_{\mathbf{W},\mathbf{S}}(\{1\},[2:M+1])$. 
By~\eqref{eq:JointPMF} and~\eqref{pmf:W}, we have 
\begin{equation}\label{eq:ratioPWSPW}
\frac{p_{\mathbf{W},\mathbf{S}}(\mathrm{W}^{*},\mathrm{S}^{*})}{p_{\mathbf{W}}(\mathrm{W}^{*})} = \frac{1}{\lambda_{\overline{\mathrm{S}}^{*}}}\left(\sum_{\mathrm{T}\in [\mathcal{K}\setminus \mathrm{W}^{*}]^{M}} \frac{1}{\lambda_{\overline{\mathrm{T}}}}\right)^{-1}.    
\end{equation}
For any given $\mathrm{W}^{*}\in \mathcal{K}$, it is easy to see that~\eqref{eq:ratioPWSPW} is minimized for $\mathrm{S}^{*}\in [\mathcal{K}\setminus \mathrm{W}^{*}]^{M}$ such that $\lambda_{\overline{\mathrm{S}}^{*}}$ is maximum, or equivalently, $\lambda_{\mathrm{S}^{*}} := \sum_{i\in \mathrm{S}^{*}} \lambda_i$ is minimum. 
For any given $\mathrm{W}^{*}$, we can determine $\mathrm{S}^{*}$ that minimizes $\lambda_{\mathrm{S}^{*}}$ as follows. 
Recall that ${\lambda_1\geq \lambda_2\geq \dots\geq \lambda_K}$ by assumption.
We consider the following two cases separately:  
(i)~${\mathrm{W}^{*}\in [1:K-M]}$, and 
(ii)~${\mathrm{W}^{*}\in [K-M+1:K]}$. 
In the case~(i), $\lambda_{\mathrm{S}^{*}}$ is minimized for ${\mathrm{S}^{*} = [K-M+1:K]}$. 
This is because the sum of the last $M$ $\lambda_i$'s yields the minimum sum over all $M$-subsets of $\{\lambda_i: i\in \mathcal{K}\}$.
In the case~(ii), $\lambda_{\mathrm{S}^{*}}$ is minimized for ${\mathrm{S}^{*} = [K-M:K]\setminus \mathrm{W}^{*}}$.
This is because $\mathrm{W}^{*}$ is one of the last $M+1$ indices in $\mathcal{K}$, and the $M$-subset $\mathrm{S}^{*}$ cannot contain $\mathrm{W}^{*}$. 
According to these results, we can rewrite the minimization problem in~\eqref{eq:search_reduced} as 
\begin{align}\label{eq:search_modified}
& \min\Biggl\{\frac{p_{\mathbf{W},\mathbf{S}}(\{1\},[2:M+1])}{p_{\mathbf{W}}(\{1\})},\nonumber \\
& \quad \quad \hspace{0.2cm} \min_{i\in [1:K-M]} 
\frac{p_{\mathbf{W},\mathbf{S}}(\{i\},[K-M+1:K])}{p_{\mathbf{W}}(\{i\})},\nonumber\\
& \quad \hspace{0.525cm} \min_{i\in [K-M+1:K]} 
\frac{p_{\mathbf{W},\mathbf{S}}(\{i\},[K-M:K]\setminus \{i\})}{p_{\mathbf{W}}(\{i\})}\Biggr\}.
\end{align}

\begin{lemma}\label{lem:intermediate}
For any popularity profile $(\lambda_1,\dots,\lambda_K)$ such that ${\lambda_1\geq \lambda_2\geq \cdots\geq \lambda_K>0}$, it holds that
\begin{align}\label{eq:intermediate}
& \min_{i\in [1:K-M]} 
\frac{p_{\mathbf{W},\mathbf{S}}(\{i\},[K-M+1:K])}{p_{\mathbf{W}}(\{i\})} \nonumber \\ 
& \quad \hspace{0.125cm} = \frac{p_{\mathbf{W},\mathbf{S}}(\{K-M\},[K-M+1:K])}{p_{\mathbf{W}}(\{K-M\})}.     
\end{align}
\end{lemma}

By the result of Lemma~\ref{lem:intermediate},
the minimization problem in~\eqref{eq:search_modified} can be simplified further as
\begin{align*}
& \min\Biggl\{\frac{p_{\mathbf{W},\mathbf{S}}(\{1\},[2:M+1])}{p_{\mathbf{W}}(\{1\})}, \\
& \quad \hspace{0.525cm} \min_{i\in [K-M:K]} 
\frac{p_{\mathbf{W},\mathbf{S}}(\{i\},[K-M:K]\setminus \{i\})}{p_{\mathbf{W}}(\{i\})}\Biggr\},
\end{align*} 
or equivalently, 
\begin{align}\label{eq:new_search_modified}
& \min_{i\in [K-M:K]}\Biggl\{\frac{p_{\mathbf{W},\mathbf{S}}(\{1\},[2:M+1])}{p_{\mathbf{W}}(\{1\})},\nonumber \\
& \quad \quad \quad \quad \hspace{0.525cm} 
\frac{p_{\mathbf{W},\mathbf{S}}(\{i\},[K-M:K]\setminus \{i\})}{p_{\mathbf{W}}(\{i\})}\Biggr\}.
\end{align}

Since~\eqref{eq:search_reduced},~\eqref{eq:search_modified}, and~\eqref{eq:new_search_modified} are equal, and~\eqref{ineq:param} is equal to~\eqref{eq:search_reduced} times  $p_{\mathbf{W}}(\{1\})/p_{\mathbf{W},\mathbf{S}}(\{1\},[2:M+1])$, then it follows that~\eqref{ineq:param} is equal to 
\begin{equation*}
\min_{i\in [K-M:K]}\left\{1,\frac{p_{\mathbf{W},\mathbf{S}}(\{i\},[K-M:K]\setminus \{i\})p_{\mathbf{W}}(\{1\}) }{p_{\mathbf{W},\mathbf{S}}(\{1\},[2:M+1])p_{\mathbf{W}}(\{i\})}\right\},    
\end{equation*} which is the same as~\eqref{eq:thm1_parameter}, as was to be shown.

\section{Simulations}
In this section, we compare the rate of the RCS scheme and that of the MDS Code scheme of~\cite{KGHERS2020}, with respect to the capacity upper bound $R_{\text{UB}}$ (see~\eqref{eq:RUB}). 
In the following, we denote the rates of the RCS and MDS Code schemes by $R_{\text{RCS}}$ and $R_{\text{MDS}}$, respectively.
Note that $R_{\text{RCS}}=R_{\text{LB}}$  (see~\eqref{eq:thm1_rate}), and $R_{\text{MDS}}=1/(K-M)$.


In practice, the popularity profile depends mostly on the type of content as well as the server's workload; however, it is generally agreed that the Zipf, Gamma, and Weibull distributions are appropriate models for the popularity profile~\cite{CKRAM2009,BCFPS99,CDL2008}. 
Motivated by this, in our simulations we have considered popularity profiles generated according to each of these distributions. 
In addition, we consider very small values of $M$, particularly, $M=1$, $2$, and $3$, which are of significant practical importance.




\begin{figure}[t]
\includegraphics[width=0.52\textwidth]{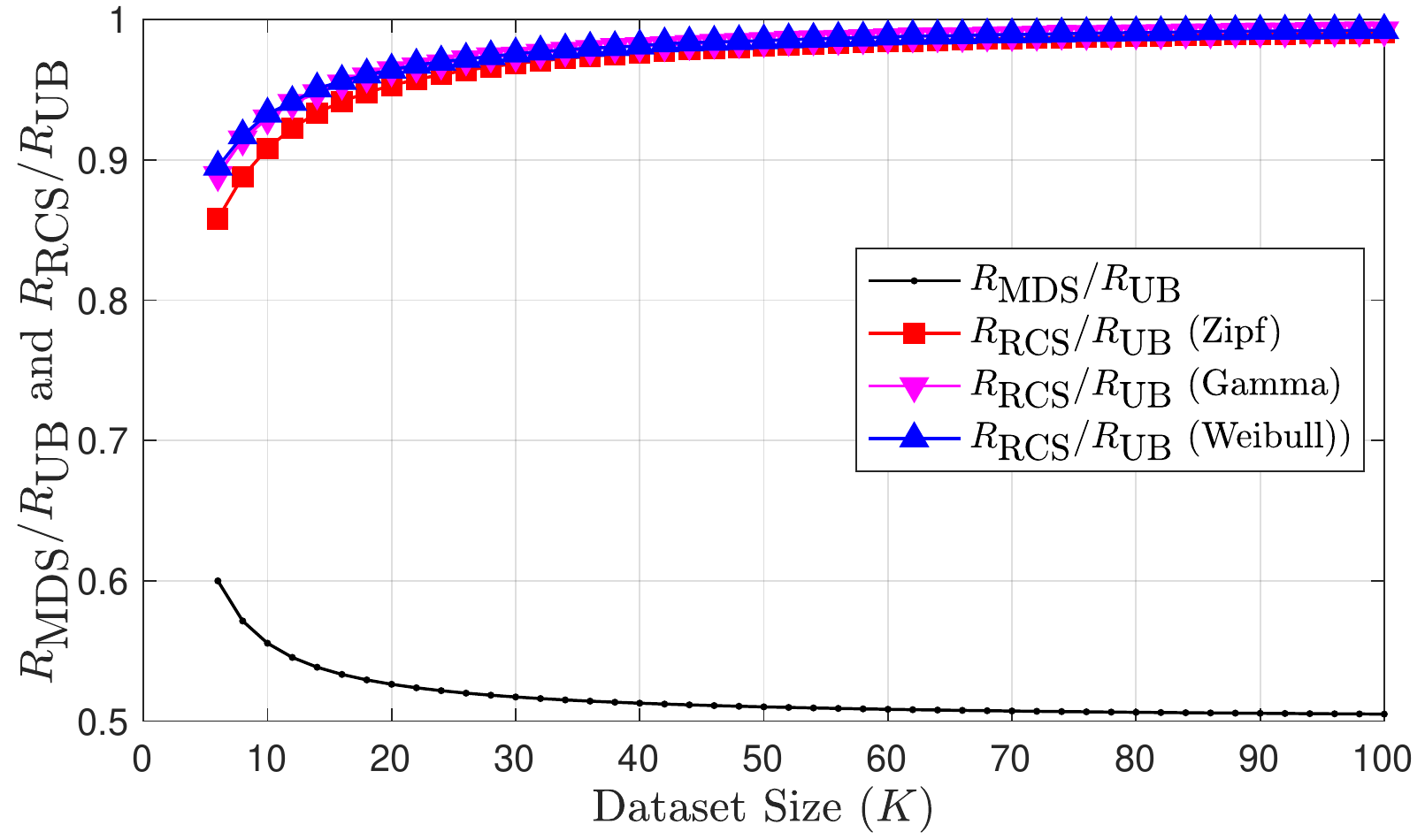}
\centering
\caption{The ratios $R_{\text{RCS}}/R_{\text{UB}}$ and $ R_{\text{MDS}}/R_{\text{UB}}$ versus $K$, for $M=1$ and different models for the popularity profile.}\label{fig:plot1}
\end{figure}

Fig.~\ref{fig:plot1} depicts the ratios $R_{\text{RCS}}/R_{\text{UB}}$ and $R_{\text{MDS}}/R_{\text{UB}}$, for $M=1$ and different $K$, where  $\lambda_1,\dots,\lambda_K$ are sampled independently from each of the following distributions: 
(i) Zipf with parameters $N=100$ and $s=1$, 
(ii) Gamma with shape and scale parameters $0.62$ and $31.22$, respectively, and 
(iii) Weibull with shape and scale parameters $0.79$ and $16.80$, respectively.
(These parameters were chosen such that all three distributions have the same mean and the same variance.) 
For each $K$ and each distribution being considered, the ratio $R_{\text{RCS}}/R_{\text{UB}}$ is averaged over 
$1000$ independently generated popularity profiles.
As seen in Fig.~\ref{fig:plot1}, for a fixed distribution, as $K$ increases, the ratio $R_{\text{RCS}}/R_{\text{UB}}$ approaches $1$, whereas the ratio $R_{\text{MDS}}/R_{\text{UB}}$ approaches $1/2$.


\begin{figure}[t]
\includegraphics[width=0.52\textwidth]{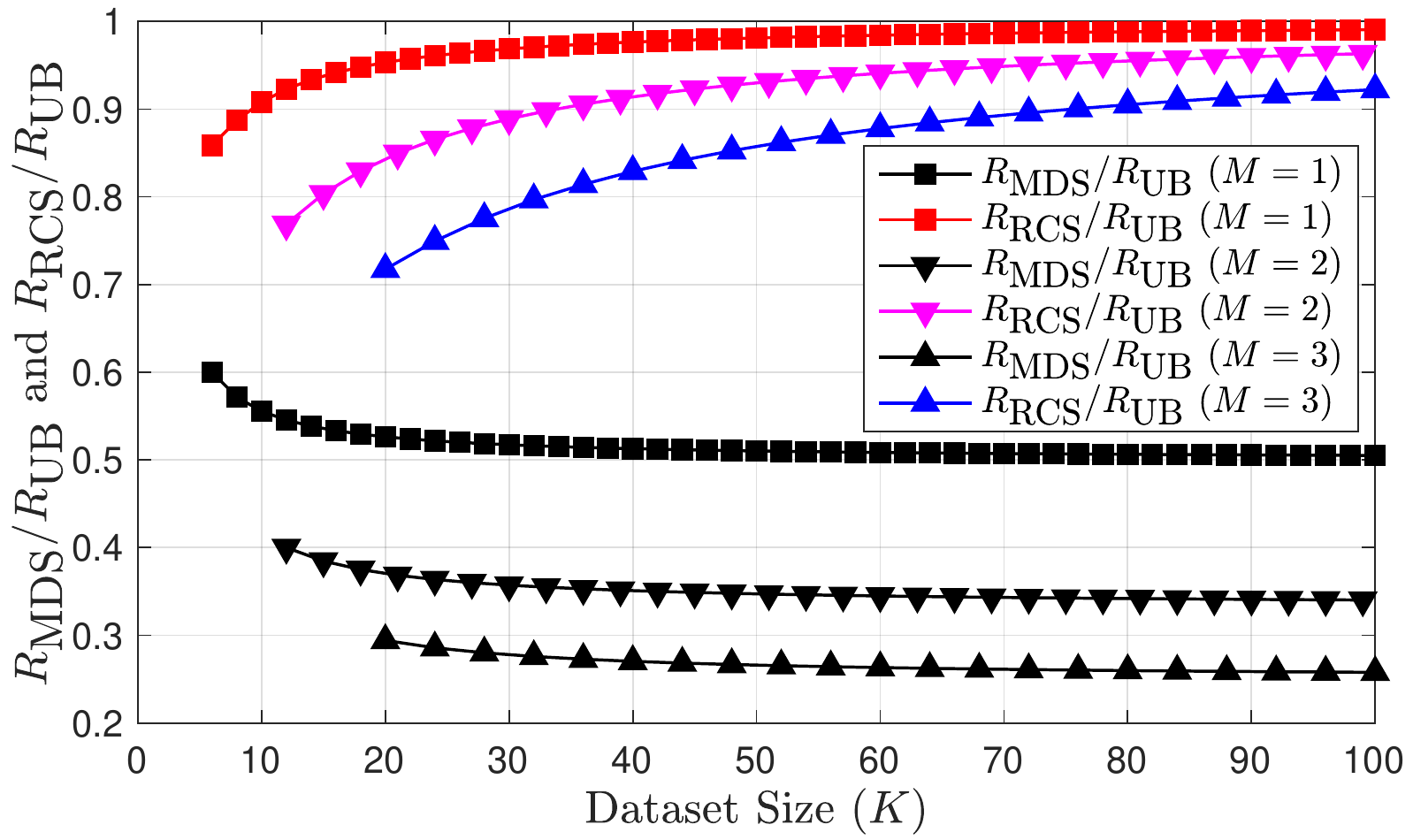}
\centering
\caption{The ratios $R_{\text{RCS}}/R_{\text{UB}}$ and $ R_{\text{MDS}}/R_{\text{UB}}$ versus $K$, for different $M$ and the Zipf model for the popularity profile.}\label{fig:plot2}
\end{figure}

Fig.~\ref{fig:plot2} depicts the ratios $R_{\text{RCS}}/R_{\text{UB}}$ and $R_{\text{MDS}}/R_{\text{UB}}$ for $M\in\{1,2,3\}$ and different $K$, where $\lambda_1,\dots,\lambda_K$ are sampled independently from the Zipf distribution with parameters $N=100$ and $s=1$.
For each pair of $M$ and $K$ being considered, the ratio $R_{\text{RCS}}/R_{\text{UB}}$ is averaged over $1000$ independently generated popularity profiles.
In Fig.~\ref{fig:plot2}, one can observe that for each $M$, as $K$ increases, the ratio $R_{\text{RCS}}/R_{\text{UB}}$ approaches $1$, while the ratio $R_{\text{MDS}}/R_{\text{UB}}$ approaches $1/(M+1)$. 
It can also be seen that for fixed $K$ (or $M$), the advantage of the RCS scheme over the MDS Code scheme is more pronounced as $M$ (or $K$) increases.


\appendix

\subsection{Proof of Lemma~\ref{lem:NC}}\label{app:NC}
The proof is by the way of contradiction. 
For an arbitrary realization ${(\mathrm{W},\mathrm{S})}$, let $\mathrm{Q}$ and $\mathrm{A}$ be the query and the corresponding answer generated by an arbitrary PA-PIR-SI protocol. 
Fix an arbitrary ${\mathrm{W}^{*}\in \mathcal{K}}$. 
Suppose that ${H(\mathbf{X}_{\mathrm{W}^{*}}| \mathbf{A}, \mathbf{Q},\mathbf{X}_{\mathrm{S}^{*}})\neq 0}$ for any ${\mathrm{S}^{*}\in [\mathcal{K}\setminus \mathrm{W}^{*}]^{M}}$, 
i.e., there does not exist any potential side information $X_{\mathrm{S}^{*}}$ given which  $\mathrm{X}_{\mathrm{W}^{*}}$ can be recovered from ${(\mathrm{A},\mathrm{Q})}$. 
Then, the server knows that the user's demand cannot be $X_{\mathrm{W}^{*}}$, i.e., 
${\mathbb{P}(\mathbf{W}=\mathrm{W}^{*}|\mathbf{Q}=\mathrm{Q}) = 0}$ 
(otherwise, if the user's demand is $X_{\mathrm{W}^{*}}$, then the decodability condition implies that the user must be able to decode $X_{\mathrm{W}^{*}}$ from ${(\mathrm{A},\mathrm{Q})}$ given their side information $\mathrm{X}_{\mathrm{S}^{*}}$ for some ${\mathrm{S}^{*}\in [\mathcal{K}\setminus \mathrm{W}^{*}]^{M}}$.) 
On the other hand, the privacy condition implies that ${\mathbb{P}(\mathbf{W}=\mathrm{W}^{*}|\mathbf{Q}=\mathrm{Q})}={\mathbb{P}(\mathbf{W}=\mathrm{W}^{*})}$. 
Thus, we must have ${\mathbb{P}(\mathbf{W}=\mathrm{W}^{*}) = 0}$. 
However, this is a contradiction because ${\mathbb{P}(\mathbf{W}=\mathrm{W}^{*})} = p_{\mathbf{W}}(\mathrm{W}^{*})\neq 0$ for any ${\mathrm{W}^{*}\in \mathcal{K}}$, noting that $p_{\mathbf{W}}(\mathrm{W}^{*})>0$ since $\lambda_1,\dots,\lambda_K>0$ by assumption (cf.~\eqref{pmf:W}).

\subsection{Proof of Lemma~\ref{lem:Claim}}\label{app:Claim}
For each $i\in [N]$, we can write
\begin{align*}
H(\mathbf{A}|\mathbf{Q},\mathbf{X}_{\mathrm{U}_i}) & \stackrel{\scriptsize{\text{(a)}}}{\geq} H(\mathbf{A}|\mathbf{Q},\mathbf{X}_{\mathrm{U}_i},\mathbf{X}_{\mathrm{S}_{i}})\\
& \stackrel{\scriptsize{\text{(b)}}}{=} H(\mathbf{A}|\mathbf{Q},\mathbf{X}_{\mathrm{U}_i},\mathbf{X}_{\mathrm{S}_{i}}) \\
& \quad + H(\mathbf{X}_{\mathrm{W}_{i}}|\mathbf{A},\mathbf{Q},\mathbf{X}_{\mathrm{U}_i},\mathbf{X}_{\mathrm{S}_{i}})\\
& \stackrel{\scriptsize{\text{(c)}}}{=} H(\mathbf{A},\mathbf{X}_{\mathrm{W}_{i}}|\mathbf{Q},\mathbf{X}_{\mathrm{U}_i},\mathbf{X}_{\mathrm{S}_{i}})\\
& \stackrel{\scriptsize{\text{(d)}}}{=} H(\mathbf{X}_{\mathrm{W}_{i}}|\mathbf{Q},\mathbf{X}_{\mathrm{U}_i},\mathbf{X}_{\mathrm{S}_{i}})\\
& \quad + H(\mathbf{A}|\mathbf{Q},\mathbf{X}_{\mathrm{U}_i},\mathbf{X}_{\mathrm{S}_{i}},\mathbf{X}_{\mathrm{W}_{i}})\\
& \stackrel{\scriptsize{\text{(e)}}}{=} H(\mathbf{X}_{\mathrm{W}_{i}}) + H(\mathbf{A}|\mathbf{Q},\mathbf{X}_{\mathrm{U}_i},\mathbf{X}_{\mathrm{S}_{i}},\mathbf{X}_{\mathrm{W}_{i}})\\
& \stackrel{\scriptsize{\text{(f)}}}{=} H(\mathbf{X}_{\mathrm{W}_{i}}) + H(\mathbf{A}|\mathbf{Q},\mathbf{X}_{\mathrm{U}_{i+1}}), 
\end{align*}
where (a) holds because conditioning does not increase the entropy; 
(b) holds because $H(\mathbf{X}_{\mathrm{W}_{i}}|\mathbf{A},\mathbf{Q},\mathbf{X}_{\mathrm{S}_{i}})=0$ (by assumption);
(c) and (d) follow from the chain rule of entropy; 
(e) holds because $\mathbf{X}_{\mathrm{W}_i}$ and $(\mathbf{Q},\mathbf{X}_{\mathrm{U}_i},\mathbf{X}_{\mathrm{S}_i})$ are independent, 
noting that $\mathrm{W}_i$ and $\mathrm{U}_i\cup \mathrm{S}_i$ are disjoint (by the choice of $\mathrm{W}_i$), ${\mathbf{X}_1,\dots,\mathbf{X}_K}$ are independent, and $\mathbf{Q}$ is independent of ${\mathbf{X}_1,\dots,\mathbf{X}_K}$ (by assumption);
and 
(f) follows because $\mathrm{U}_{i+1} = \mathrm{U}_i\cup (\mathrm{W}_i\cup \mathrm{S}_i)$ (by definition). 

\subsection{Proof of Lemma~\ref{lem:NCQ}}\label{app:NCQ}
Consider an arbitrary partition $\mathrm{Q}\in \mathcal{Q}$. 
Fix an arbitrary part $\mathrm{Q}_{i}$ in $\mathrm{Q}$, and let $\mathrm{W}_{i}$ be an arbitrary index in the part $\mathrm{Q}_{i}$, and let $\mathrm{S}_{i} = \mathrm{Q}_{i}\setminus \mathrm{W}_{i}$.
For the privacy condition to be satisfied, we require that
\begin{align}\label{eq:privacyprime}
    \mathbb{P}(\mathbf{W}=\mathrm{W}_{i}|\mathbf{Q}=\mathrm{Q}) = \mathbb{P}(\mathbf{W}=\mathrm{W}_{i}).
\end{align}
Given $\mathbf{Q}=\mathrm{Q}$, the event $\mathbf{W}=\mathrm{W}_{i}$ implies the event $\mathbf{S}=\mathrm{S}_{i}$; otherwise, if $\mathbf{S}\neq\mathrm{S}_{i}$, then the server knows that $\mathbf{W}\neq \mathrm{W}_{i}$. 
(This is because given the query $\mathrm{Q}$ and its corresponding answer $\mathrm{A}$, the message $X_{\mathrm{W}_{i}}$ can only be recovered if the messages $X_{\mathrm{S}_{i}}$ are known.)  
By applying Bayes' rule, we have
\begin{align}\label{eq:WSprime}
    & \mathbb{P}(\mathbf{W}=\mathrm{W}_{i}|\mathbf{Q}=\mathrm{Q}) \nonumber \\
    & = \frac{\mathbb{P}(\mathbf{Q} = \mathrm{Q}| \mathbf{W} = \mathrm{W}_{i}, \mathbf{S} = \mathrm{S}_{i})p_{\mathbf{W},\mathbf{S}}(\mathrm{W}_{i},\mathrm{S}_{i})}{\mathbb{P}(\mathbf{Q}=\mathrm{Q})}
\end{align}
Let $L$ be the number of ways to partition $K-(M+1)$ distinct elements into $N-1$ parts, each of size $M+1$. 
Then, for the privacy condition to be satisfied, it must hold that
\begin{align}
    \mathbb{P}(\mathbf{Q} = \mathrm{Q}) & \stackrel{\scriptsize{\text{(a)}}}{=} \frac{\mathbb{P}(\mathbf{Q} = \mathrm{Q}| \mathbf{W} = \mathrm{W}_{i}, \mathbf{S} = \mathrm{S}_{i})p_{\mathbf{W},\mathbf{S}}(\mathrm{W}_{i},\mathrm{S}_{i})}{p_{\mathbf{W}}(\mathrm{W}_{i})} \nonumber \\
    & \stackrel{\scriptsize{\text{(b)}}}{=} \frac{\Gamma_{\mathrm{W}_{i},\mathrm{S}_{i}} \times \frac{1}{N}\times \frac{1}{L} \times p_{\mathbf{W},\mathbf{S}}(\mathrm{W}_{i},\mathrm{S}_{i})}{p_{\mathbf{W}}(\mathrm{W}_{i})}\label{eq:forallWS}
\end{align}
where (a) follows from combining~\eqref{eq:privacyprime} and~\eqref{eq:WSprime} and rearranging terms, and 
(b) holds because given $\mathbf{W}=\mathrm{W}_{i}$ and $\mathbf{S}=\mathrm{S}_{i}$, 
the user first selects Scheme~I with probability $\Gamma_{\mathrm{W}_{i},\mathrm{S}_{i}}$; 
then, the user assigns all $M+1$ indices in $\mathrm{W}_{i}\cup \mathrm{S}_{i}$ to a part chosen uniformly at random among all $N$ parts; 
and finally, the user chooses one of the $L$ possible ways to partition the remaining $K-(M+1)$ indices in the remaining $N-1$ parts, also uniformly at random.

Fix an arbitrary part $\mathrm{Q}_{j}$ in $\mathrm{Q}$, and let $\mathrm{W}_{j}$ be an arbitrary index in the part $\mathrm{Q}_{j}$, and let $\mathrm{S}_{j} = \mathrm{Q}_{j}\setminus \mathrm{W}_{j}$.
By the same arguments as in~\eqref{eq:privacyprime}-\eqref{eq:forallWS}, for the privacy condition to be satisfied, we must have
\begin{equation}\label{eq:forallWS2}
\mathbb{P}(\mathbf{Q} = \mathrm{Q}) = \frac{\Gamma_{\mathrm{W}_j,\mathrm{S}_j} \times \frac{1}{N}\times \frac{1}{L} \times p_{\mathbf{W},\mathbf{S}}(\mathrm{W}_j,\mathrm{S}_j)}{p_{\mathbf{W}}(\mathrm{W}_j)}.
\end{equation}
By combining~\eqref{eq:forallWS} and~\eqref{eq:forallWS2}, it follows that the privacy condition is satisfied so long as
\begin{align*}
& \frac{\Gamma_{\mathrm{W}_{i},\mathrm{S}_{i}} \times \frac{1}{N}\times \frac{1}{L} \times p_{\mathbf{W},\mathbf{S}}(\mathrm{W}_{i},\mathrm{S}_{i})}{p_{\mathbf{W}}(\mathrm{W}_{i})} \nonumber \\ 
& \quad = \frac{\Gamma_{\mathrm{W}_j,\mathrm{S}_j} \times \frac{1}{N}\times \frac{1}{L} \times p_{\mathbf{W},\mathbf{S}}(\mathrm{W}_j,\mathrm{S}_j)}{p_{\mathbf{W}}(\mathrm{W}_j)},
\end{align*}
or equivalently,
\begin{equation*} 
 \Gamma_{\mathrm{W}_j,\mathrm{S}_j}= \Gamma_{\mathrm{W}_{i},\mathrm{S}_{i}} \frac{p_{\mathbf{W},\mathbf{S}}(\mathrm{W}_{i},\mathrm{S}_{i})p_{\mathbf{W}}(\mathrm{W}_j) }{p_{\mathbf{W},\mathbf{S}}(\mathrm{W}_j,\mathrm{S}_j)p_{\mathbf{W}}(\mathrm{W}_{i})},
\end{equation*} 
as was to be shown.

\subsection{Proof of Lemma~\ref{lem:intermediate}}\label{app:intermediate}
Taking $\mathrm{W}^{*} = \{i\}$ and $\mathrm{S}^{*} = [K-M+1:K]$ in~\eqref{eq:ratioPWSPW}, 
\begin{align}\label{eq:RHS}
& \frac{p_{\mathbf{W},\mathbf{S}}(\{i\},[K-M+1:K])}{p_{\mathbf{W}}(\{i\})} \nonumber\\
& \quad = \frac{1}{\lambda_{\overline{[K-M+1:K]}}}\left(\sum_{\mathrm{T}\in [\mathcal{K}\setminus \{i\}]^{M}} \frac{1}{\lambda_{\overline{\mathrm{T}}}}\right)^{-1}.    
\end{align}
Fix arbitrary $i_1,i_2\in [K-M]$ such that $i_1\leq i_2$.  
To show~\eqref{eq:intermediate}, it suffices to show that  
\begin{equation}\label{eq:sufficient}
\sum_{\mathrm{T}\in [\mathcal{K}\setminus \{i_1\}]^{M}} \frac{1}{\lambda_{\overline{\mathrm{T}}}}\leq \sum_{\mathrm{T}\in [\mathcal{K}\setminus \{i_2\}]^{M}} \frac{1}{\lambda_{\overline{\mathrm{T}}}}.    
\end{equation}
Let $\mathcal{T}_1$ (or $\mathcal{T}_2$) be the set of all $M$-subsets of $\mathcal{K}\setminus \{i_1\}$ (or $\mathcal{K}\setminus \{i_2\}$) that contain $i_2$ (or $i_1$). 
Using these notations, it is easy to see that~\eqref{eq:sufficient} can be rewritten as 
\begin{equation}\label{eq:sufficient2}
\sum_{\mathrm{T}\in \mathcal{T}_1} \frac{1}{\lambda_{\overline{\mathrm{T}}}}\leq \sum_{\mathrm{T}\in \mathcal{T}_2} \frac{1}{\lambda_{\overline{\mathrm{T}}}}.    
\end{equation}
Let ${R\triangleq |\mathcal{T}_1|}$, and let $\mathcal{T}_1 = \{\mathrm{T}_1,\dots,\mathrm{T}_R\}$.
It is easy to see that 
$\mathcal{T}_2 = \{(\mathrm{T}_1\cup\{i_1\})\setminus \{i_2\},\dots,(\mathrm{T}_R\cup\{i_1\})\setminus \{i_2\}\}$. 
Then, we can rewrite~\eqref{eq:sufficient2} as
\begin{equation}\label{eq:sufficient3}
\sum_{j\in [R]} \frac{1}{\lambda_{\overline{\mathrm{T}}_j}}\leq \sum_{j\in [R]} \frac{1}{\lambda_{\overline{(\mathrm{T}_j\cup \{i_1\})\setminus \{i_2\}}}}.     
\end{equation}
To prove~\eqref{eq:intermediate}, we thus need to show that~\eqref{eq:sufficient3} is satisfied.
It is easy to verify that 
${\lambda_{\overline{(\mathrm{T}_j\cup \{i_1\})\setminus \{i_2\}}} = \lambda_{\overline{\mathrm{T}}_j}+\lambda_{i_2}-\lambda_{i_1}}$.
Note that
${\lambda_{\overline{\mathrm{T}}_j} - \lambda_{\overline{(\mathrm{T}_j\cup \{i_1\})\setminus \{i_2\}}}} = {\lambda_{i_1}-\lambda_{i_2} \geq 0}$. 
This is because ${i_1\leq i_2}$, and hence, $\lambda_{i_1}\geq \lambda_{i_2}$ (by assumption).
Thus, for all ${j\in [R]}$, ${\lambda_{\overline{\mathrm{T}}_j} \geq \lambda_{\overline{(\mathrm{T}_j\cup \{i_1\})\setminus \{i_2\}}}}$, or equivalently, 
\begin{equation}\label{eq:final}
\frac{1}{\lambda_{\overline{\mathrm{T}}_j}} \leq \frac{1}{\lambda_{\overline{(\mathrm{T}_j\cup \{i_1\})\setminus \{i_2\}}}}.    
\end{equation}
Summing both sides of~\eqref{eq:final} over all $j\in [R]$, we arrive at~\eqref{eq:sufficient3}, as was to be shown.
This completes the proof.  

\bibliographystyle{IEEEtran}
\bibliography{PIR_PC_Refs}

\end{document}